\documentclass[10pt,journal,cspaper,compsoc]{IEEEtran}
\usepackage{graphicx}
\usepackage{caption}
\usepackage{amssymb, amsmath}
\usepackage{epstopdf}
\usepackage{url}
\usepackage{caption}
\usepackage{subcaption}
\usepackage{multirow}
\usepackage{lipsum}
\usepackage{tabularx}
\DeclareMathOperator*{\argmin}{\arg\!\min}
\usepackage{makecell}
\usepackage{marginnote}
\usepackage{graphics}
\usepackage{times}
\usepackage{cite}

\usepackage{enumitem}
\usepackage[T1]{fontenc}
\usepackage[font=small,labelfont=bf,tableposition=top]{caption}

\def \sys {{CheckInside}}

\begin{document}
\title{A Fine-grained Indoor Location-based Social  Network}
\author{Moustafa ~Elhamshary,~\IEEEmembership{Student~Member,~IEEE,}
        Anas ~ Basalamah,~\IEEEmembership{Member,~IEEE,}
        and~Moustafa~Youssef,~\IEEEmembership{Senior~Member,~IEEE}
\IEEEcompsocitemizethanks{\IEEEcompsocthanksitem Moustafa Elhamshary, and Moustafa Youssef  are with the Wireless Research Center, Egypt-Japan Univ. for Sci. and Tech. (E-JUST),  Egypt.  Moustafa Elhamshary is currently  a research fellow at the Grad. Sch. of Info. Sc. and Tech., Osaka Univ., Japan. Moustafa Youssef is currently on sabbatical from Alexandria Univ., Egypt.\protect\\
E-mail: \{mostafaelhamshary; moustafa.youssef\}@ejust.edu.eg

\IEEEcompsocthanksitem Anas Basalmah is  with the Comp. Eng. Dept., KACST GIS Tech. Innov. Ctr.
Umm Al-Qura Univ., Saudi Arabia.\protect\\
E-mail: abasalamah@gistic.org}

\thanks{An earlier  version of this paper appeared in the proc. of The 2014 ACM Int. Joint Conf. on Perv. and Ubiq. Comp. (UbiComp 2014) \cite{elhamshary2014checkinside}.}
}

\markboth{IEEE Transaction on mobile computing,~Vol.~x, No.~x, August~2016}
{ \MakeLowercase{\textit{et al.}}: An Indoor Location-based Social Network }
\IEEEcompsoctitleabstractindextext{
\begin{abstract}

Existing  Location-based social networks (LBSNs), e.g., Foursquare, depend  mainly on GPS or cellular-based localization to infer users' locations. However, GPS is unavailable indoors and cellular-based localization provides coarse-grained accuracy. This limits the  accuracy of current LBSNs in indoor environments, where people spend 89\% of their time. This in turn affects the user experience, in terms of the accuracy of the ranked list of venues, especially for the small screens
of mobile devices; misses business opportunities; and
leads to reduced venues coverage.

In this paper, we present CheckInside: a system that can provide a fine-grained indoor location-based social network. CheckInside leverages the crowd-sensed data collected from users' mobile devices during the check-in operation and knowledge extracted from current LBSNs to associate a place with a logical name and a semantic fingerprint. This semantic fingerprint is used to obtain  a more accurate list of nearby places as well as to automatically detect new places with similar signature.  A novel algorithm for  detecting fake check-ins and inferring a semantically-enriched floorplan  is proposed  as well as an algorithm for enhancing the system performance based on the user implicit feedback. Furthermore, CheckInside encompasses a coverage extender module  to automatically predict names  of new venues  increasing  the coverage of current LBSNs.

Experimental evaluation of CheckInside in four malls over the course of six weeks with 20 participants shows that it can infer the actual user place within the top five venues 99\% of the time. This is compared to 17\% only in the case of current LBSNs. In addition, it increases the coverage of existing LBSNs by more than   37\%.
\end{abstract}}

\maketitle
\IEEEdisplaynotcompsoctitleabstractindextext
\IEEEpeerreviewmaketitle
\section{Introduction}
\IEEEPARstart{S}ocial networking applications, e.g., Facebook, have become one of the most important web services that provide Internet-based platforms for users to interact with other people that are socially-relevant to them. With the advances in location determination technologies, the flourishing of GPS-equipped mobile devices, and the development of wireless Internet connectivity during the last decade; location-based social networks (e.g., Foursquare, Facebook Places, etc) started to emerge. Such LBSNs allow users to share their location information with other people in their social structure \cite{zheng2011location}. In addition, they provide services with spatial relevance to the users such as finding interesting places within a certain geographical area. Moreover, LBSNs provide businesses opportunities for better user reach, including location-based ads and location-based business analytics. This leads to a wide interest in such networks both from academia and industry with companies such as Foursquare reporting nearly 55 million users with over 7 billions check-ins and millions more check-ins every day\footnote{\url{https://foursquare.com/about} (last accessed Sep. 2015.)}.

The main interaction among users in LBSNs is location sharing through the notion of check-in where users  voluntarily share their locations with their peers. During the check-in operation, the user is presented with a ranked list of nearby venues to choose her current location. With
the limited screen size of mobile phones, accurate ranking
of location-based query results becomes crucial as the user
would find it hard to scroll beyond the top few results. A number of approaches have been proposed in literature to tackle  the venues ranking problem in LBSNs. These approaches either rely on experts to evaluate  places, rely on the review of all users that visited these places previously, rank places based on the closest distance to the estimated user location, or based on places popularity \cite{shankar2012crowds,yelp}. Regardless of the ranking algorithm used,  places ranking  usually depends on the accurate localization of the phone user for better efficiency and accuracy in location queries. However, traditional LBSNs depend on the GPS and/or network-based localization techniques. Consequently, current LBSNs provide reasonable accuracy \textbf{\emph{only}} for outdoor environments or entire buildings.

On the other hand, in indoor environments, GPS is unreliable  and the accuracy of cellular-based approaches range from a few hundred meters to kilometers.
 Even when WiFi is turned on (e.g., using Google MyLocation), our experiments below show that the median distance error in estimating the actual venue location is 84m, which is still coarse-grained for indoor environments.  Such inaccuracy leads to a worse user experience, which in turn is reflected on the accuracy of the collected data and  the business value.  With the fact that users spend about 89\% of their time indoors \cite{klepeis2001national}, this sparks the need for a new LBSN that can work well in indoor environments.

\setlength{\belowcaptionskip}{-6pt}

 Directly extending current LBSNs to use an accurate indoor location determination technique from literature does not solve the problem (as we quantify in the evaluation section) since there are a number of challenges that need to be addressed in order to have a truly indoor LBSN. Specifically, all indoor localization techniques that leverage smartphones sensors, including WiFi, have an average localization error in the range of few meters. This error in localization can lead to placing the user on the other side of the wall in a completely different venue \cite{azizyan2009surroundsense}.
Moreover, as LBSNs are organic systems which are based on users' contribution,  their  data are susceptible to  some noise in the form of  incorrect check-ins.  These  errors lead to problems in venue ranking and labelling. Furthermore, the system needs to be energy-efficient  to avoid phone battery drainage. Finally, and \textbf{most importantly}, an indoor LBSN should learn the labels of indoor locations automatically to answer nearest-location queries efficiently and accurately. This cannot be done manually for scalability reasons and due to the inaccuracies of user check-ins and location.

In this paper, we introduce CheckInside: a fine-grained indoor LBSN that combines physical and logical localization techniques to address the above challenges and identify the user actual place accurately. The core idea is to link crowd-sensed data collected from users' smart phones  during the check-in operation or opportunistically; with the available venues information  retrieved from the traditional LBSNs. When the user performs a check-in operation, multi-modal sensor information (e.g., inaccurate indoor location, opportunistic images and audio samples, etc), are processed by the CheckInside server to construct a sensor-based fingerprint for the current user location. This fingerprint is filtered and matched against different venues fingerprints stored in the CheckInside venues database, which is constructed from the current information in traditional LBSNs and information extracted  from previous check-ins in a crowd-sensing approach. The  candidate venues achieving the closest matches with the user current place are  then returned and displayed as a ranked list to the user. The venue selected by the user, to check-in at, is implicitly used to label the location, update the venues fingerprint database, as well as provide a dynamic feedback on the quality of the different sensors. All sensors used by CheckInside either have a low-energy profile, are already used for other purposes, or are explicitly used by the user. Hence, CheckInside is energy-efficient.
To further address the inherent inaccuracy in indoor localization and fake check-ins taking place outside the actual venue, CheckInside employs a novel outlier detection technique to  distinguish  fake check-ins from correct ones. This allows CheckInside to determine the true fingerprint of a particular venue, and consequently using  only correct check-ins  for  floorplan semantic labelling and for assessing the weights of the different sensors in the feedback module. In addition, CheckInside can extend the coverage of current LBSNs using  a coverage extender module that can predict the names of uncovered venues. \\
 We implemented CheckInside on Android phones and evaluated it in four malls with 711 stores over six weeks with 20 different users. Our results show that it can provide the actual venue within the top five list in 99\% of the cases as compared to 17\% only in Foursquare. In addition, CheckInside can accurately  detect new venues,  increasing the coverage of current LBSN by more than 37\%.
Our main contributions are summarized as follow:
\begin{itemize}
\item We conduct a study to assess the performance of current LBSNs in indoor areas. Our study reveals  interesting findings regarding  the limitations of current  LBSNs in terms of coverage  and quality of the ranked venues list (Sec.~\ref{sec:study}).
\item We present the architecture and details of the CheckInside system as a fine-grained indoor LBSN that can address the limitations of the current LBSNs, provide semantic-rich floorplans, as well as increase the venues coverage of current LBSNs (Sec.~\ref{sec:InCheckIn} and \ref{sec:CheckInsidearc}).

\item We implement the CheckInside  system and thoroughly evaluate its performance (Sec.~\ref{sec:eval}).

\end{itemize}

Finally, sections~\ref{sec:related}, \ref{sec:disc}, and \ref{sec:conclude} discuss related work, highlight system limitations, and conclude the paper respectively.

\begin{table}
\begin{center}
\resizebox{0.45\textwidth}{!}{
    \begin{tabular}{|l|p{6cm}|}
    \hline
     Category & Sub-categories\\ \cline{1-2}
  \bf Food \& Restaurants & restaurant, cafe, dessert shop, ice-cream shop, bakery\\ \cline{1-2}
    \bf Clothing \& Fashion  &clothing store, accessories store,  shoe store, cosmetic store, jewelry store \\ \cline{1-2}
   \bf Entertainment \& Arts & cinema, theater, gym, gaming room, pool hall\\ \cline{1-2}
   \bf Others &book store, bar, salon, high-tech outlet, grocery store, department store, supermarket\\\cline{1-2}
       \end{tabular}
}
\end{center}
\caption{Venues categories.}
\label{tab:cat}
\end{table}

\section{Study of the Limitations of Current LBSNs for Indoor Environments}
\label{sec:study}
To motivate our work, we conducted a study to quantify the limitations of traditional LBSNs for indoor location based services as well as  illustrate the characteristics of the check-in data at these LBSNs. We use Foursquare in our study. We investigate the limitations of Foursquare in terms of two main factors: coverage (the number of indoor venues covered by a LBSN to the total number of venues available) and quality of location information (ranking and distance error of the actual venue in the list of nearby venues). Moreover, we quantify  how frequently certain stores exist in more than one shopping mall (i.e., chain of stores  belonging to the same brand) which can be leveraged to increase LBSNs venues coverage ratio. To perform this study, we developed an Android application that uses the Foursquare API to perform  check-ins at venues visited by  participants. When a contributor issues a check-in query, the application consults Foursquare to retrieve the list of nearby venues that satisfy the user query. The retrieved list along with the ground truth venue, selected by the contributor or manually entered if the actual venue is not presented in the list, are stored on the phone for later analysis.

We  surveyed 711 stores in four different malls by 20 people. Venues covered by the study are divided into four categories as shown in Table~\ref{tab:cat}. We had  two modes of operation for WiFi: on and off. When \emph{\textbf{WiFi was turned on}}, as recommended by Foursquare application, this leads to higher localization accuracy as compared to using cellular localization as quantified in the next subsections.

\subsection{Coverage Study}
\begin{figure}[!t]
    \centering
    \includegraphics[scale=0.35]{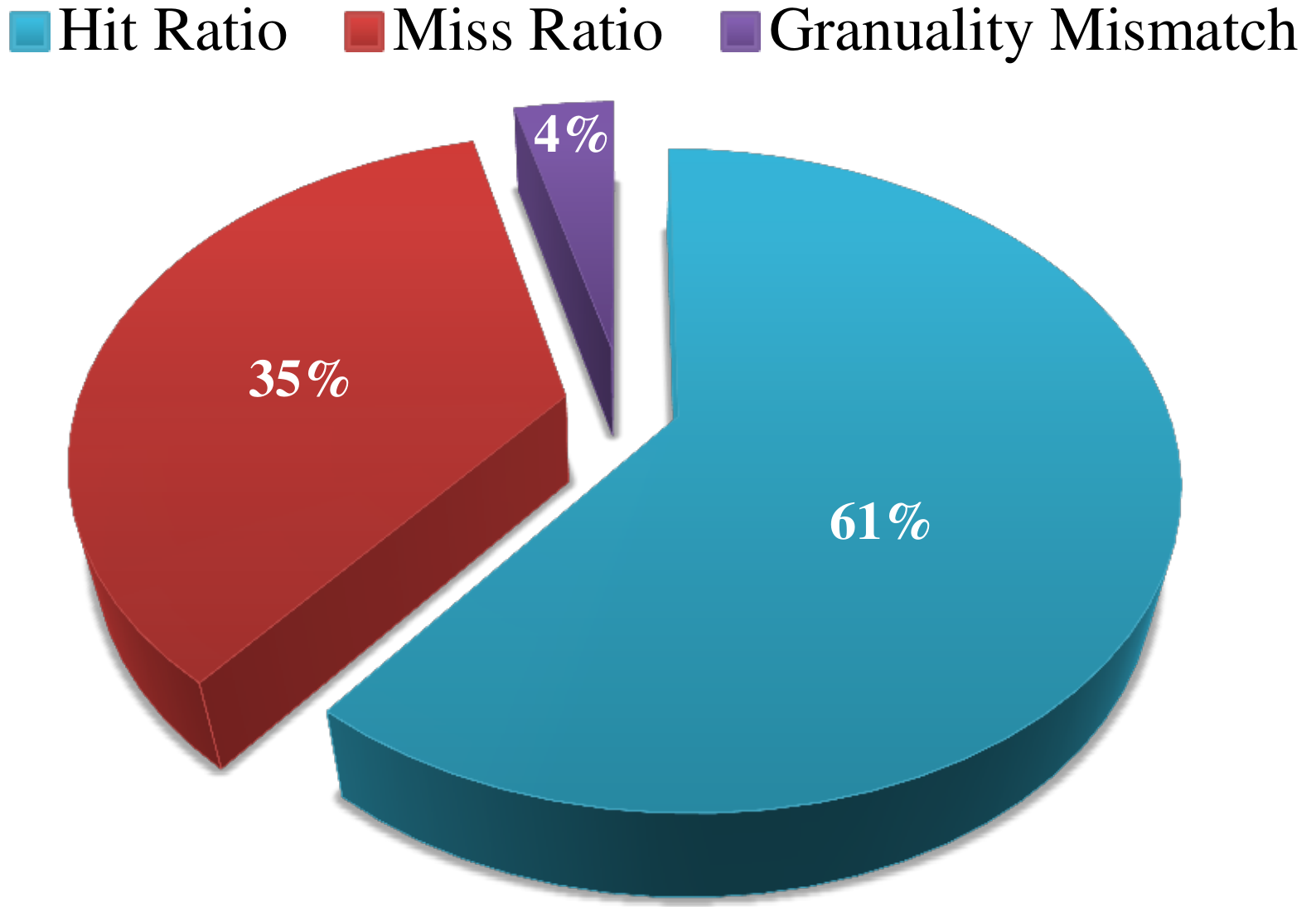}
    \caption{Venues coverage in Foursquare. "Granularity mismatch" refers to identifying a venue (e.g., a specific restaurant) by a coarse-grained label (e.g., food court).}
    \label{fig:studty_all}
\end{figure}

Coverage refers to the percentage of places that are included in the Foursquare database. Fig.~\ref{fig:studty_all} gives the overall coverage statistics and Table~\ref{tab:coverage} gives the category details. Our study shows that there are three main issues:  missed venues, granularity mismatch, and duplicate entries.

First, Foursquare misses about 39\% of venues in the four malls included in the study (not registered at all or registered at a different granularity as discussed below).

Additionally, there is a mismatch between the users' expectations of a place name and the label returned by Foursquare (\textbf{granularity mismatch issue}). For example, for some restaurants, Foursquare reports "food court" as the name of the venue (the actual venue name is not registered in the Foursquare database), which is not expressive enough for  participants about their current place, as the food court area contains a large number of venues. This contributes to 4\% of the venues in the study.

Finally, we noticed also that 8\% of the venues were registered more than once with slightly different names. We believe that the reason for this redundancy is that some users failed to find their current venue in the returned list from Foursquare due to inefficiencies in its ranking function and opted to register a new name.

 Moreover, the coverage and the granularity mismatch problems of Foursquare are much worse in non-business buildings such as educational and residential venues. For example, in our university campus, only the university name and the names of buildings are covered (e.g., no lectures halls names or department names).

\begin{table}[!t]
\label{tab:coverage}
\begin{center}
\resizebox{0.45\textwidth}{!}{
    \begin{tabular}{ |l| l | l | l |l}
    \hline
    \bf Category &\parbox[t]{1.2cm}{\#actual \\ venues}& \parbox[t]{1.3cm}{\#covered\\ venues} &\parbox[t]{1.3cm}{\ \% of \\    coverage}\\ \cline{1-4}
   \bf Food \& Restaurants & 101 & 91& 90.0\% \\ \cline{1-4}
   \bf  Clothing \& Fashion   & 374& 235 &62.8\%  \\ \cline{1-4}
   \bf  Entertainment \& Arts & 23 & 14 &  60.8\% \\ \cline{1-4}
   \bf Others & 213& 96& 45.0\%\\ \hline\hline
    \bf Total &711& 436& 61.3\%\\\cline{1-4}
    \end{tabular}
    }
\end{center}
\caption{Summary of indoor coverage for each category.}
\label{tab:coverage}
\end{table}

\begin{table}[!t]
\begin{minipage}[t]{0.485\linewidth}
\resizebox{1\textwidth}{!}{
\begin{tabular}{|p{2.5cm}|c|}
  \hline
 \bf  Avg. check-ins/venue &122.5 \\ \cline{1-2}
  \bf Avg. users/venue &57.2  \\ \cline{1-2}
  \bf  Avg. tips/venue&5.4 \\ \cline{1-2}
 \bf \ \% of check-ins done by the same user/venue&53.4\% \\ \cline{1-2}
  \end{tabular}
  }
  \vspace{0.3cm}  \caption{Average check-ins statistics per venue.}
  \label{tab:stat}
\end{minipage}
\begin{minipage}[t]{0.485\linewidth}
\tiny
\resizebox{1\textwidth}{!}{
\begin{tabular}{|l|c|}
  \hline
  Category& Percentage\\\hline
 \bf  Food \& Restaurants &88.1\% \\\hline
  \bf Clothing \& Fashion &87.2\% \\\hline
  \bf Arts \& Entertainment&56.5\% \\\hline
 \bf  Others&73.7\%\\\hline
  \hline \bf Overall&82.3\%\\\hline
  \end{tabular}}
  \vspace{0.3cm}
  \caption{Percentage of brand stores in the four malls.}
  \label{tab:catcov}
\end{minipage}
\end{table}

The coverage problem differs by category as shown in Table~\ref{tab:coverage}. It is observed that most covered venues are those where users spend a considerable amount of time like food venues. In contrast, other venues like clothing stores have high miss ratios as users in these venues are busy browsing items and may not have enough time to perform check-ins.

Table~\ref{tab:stat}  provides  statistics from Foursquare check-ins  data including:  the average  number of check-ins  performed at each venue, the average number of tips  and the average  number of users at each venue, and how frequently users repeat check-ins at the same venue. As evident from the table, venues have sufficient number of tips (5.4 in our case), which motivate us to leverage words extracted from user tips to infer the user place. Moreover, given that  on  average 53.4\% of total check-ins at all venues are repeated by the same users, the user-venue familiarity  can be harnessed  as a feature in the place inference.

Finally as illustrated in Table~\ref{tab:catcov}, the majority (about 82\%)  of shops in the four malls contained in this study are local and international brand venues that have  many branches (chain of stores) distributed across different geographical locations. This  information is valuable for CheckInside to increase  the  venues coverage (as we discuss in Sec.~\ref{sec:covext}).
\subsection{Quality Study}
 To assess the quality of location information provided by Foursquare in indoor places, our study answers two questions:
(1) What is the average error in distance between the actual venue and the top venue in the ranked list of nearby venues provided by Foursquare? and (2) What is the rank of the actual venue in the list of nearby venues?

To calculate the inter-venues distance, we have used the shortest door-to-door walking distance. Our study comprises two cases of the WiFi connectivity: on and off.

\begin{figure}[!t]
        \centering
        \begin{subfigure}[b]{0.23\textwidth}
                \includegraphics[width=\textwidth,height=2.5cm]{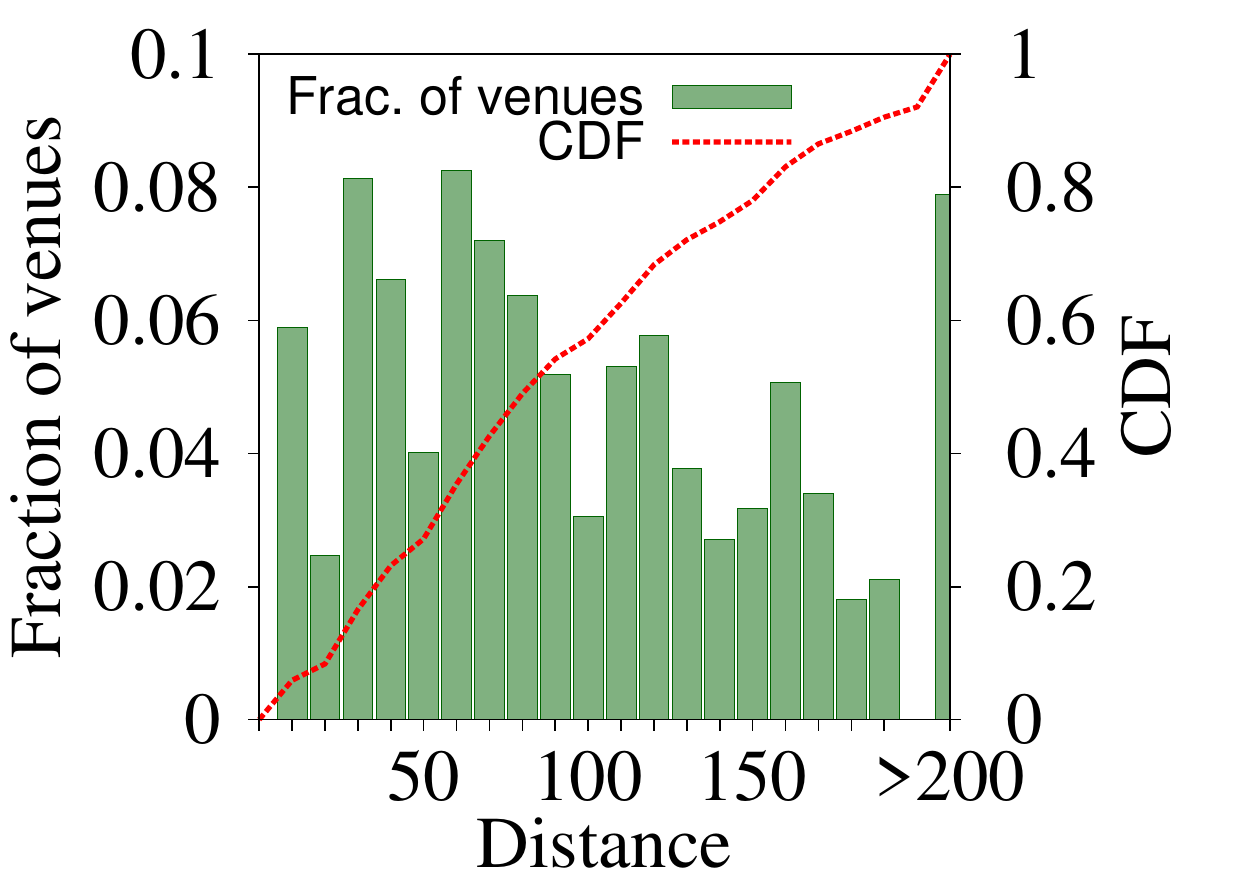}
           \caption{Distance error between the actual venue
 and first venue in the list.}
        \end{subfigure}
        \quad
                \begin{subfigure}[b]{0.23\textwidth}
                \includegraphics[width=\textwidth,height=2.5cm]{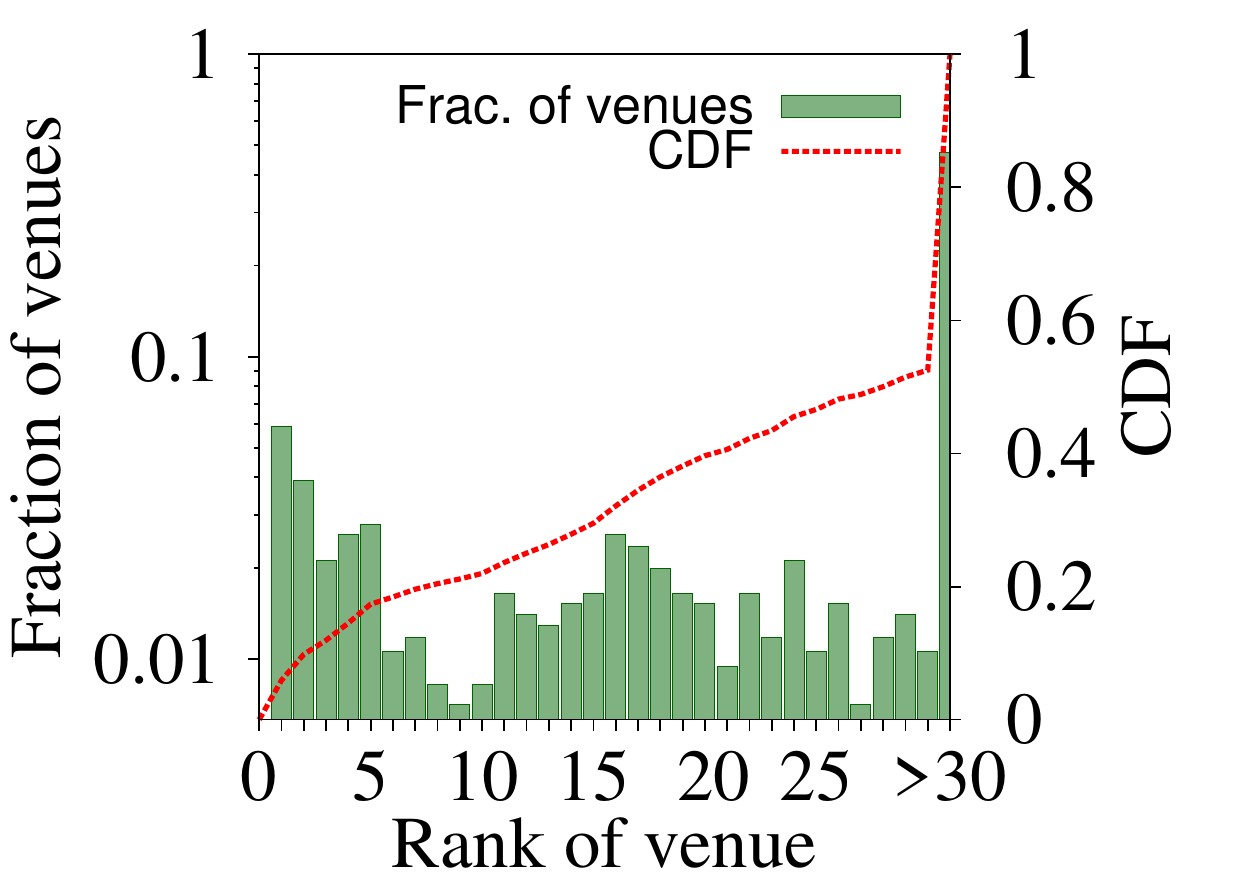}
              \caption{The rank of the actual venue in the list of nearby venues.}

        \end{subfigure}

        \caption{Quantifying the quality of the default Foursquare ranking function (WiFi on).}
\label{fig:rankdis}
\end{figure}
\textbf{WiFi turned on:}
Fig.~\ref{fig:rankdis}(a) shows that the median distance error is about 84m, which is not suitable for indoor environments. Similarly, for the second question regarding venues ranking, Fig.~\ref{fig:rankdis}(b) shows that more than 47\% of actual venues has a rank that is higher than 30 in the list returned by Foursquare. The  actual places that were not provided in the returned list are either not included  in the Foursquare databases (74\% of  cases) or covered venues that are ranked beyond  the default list size  of Foursquare (i.e., 30). In addition, we observed from the collected data that there are about 6\% of the reported venues that are outdoor venues  (even though the user was indoors).

\begin{figure}[!t]
        \centering
        \begin{subfigure}[b]{0.23\textwidth}
                \includegraphics[width=\textwidth,height=2.5cm]{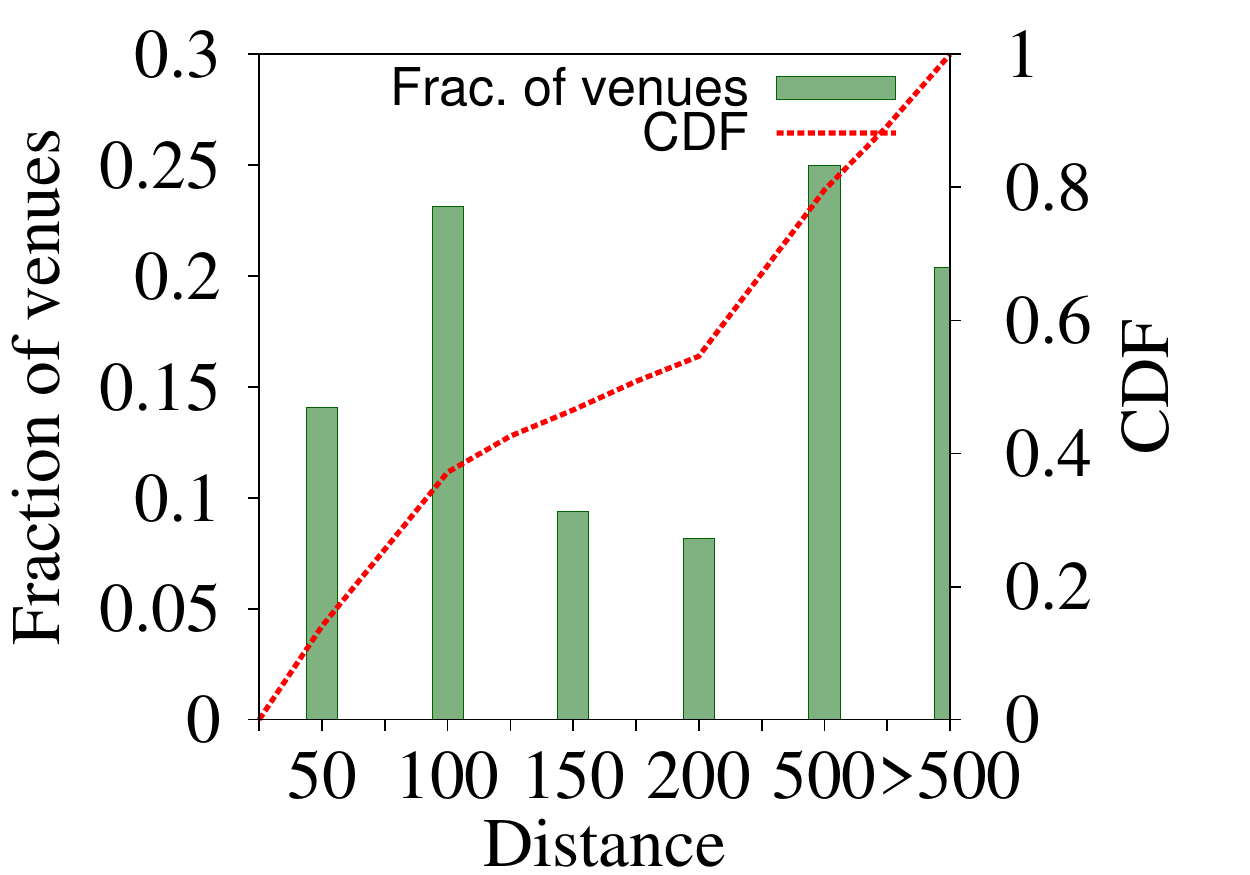}
           \caption{Distance error between the actual venue
 and first venue in the list.}
        \end{subfigure}
        \quad
                \begin{subfigure}[b]{0.23\textwidth}
                \includegraphics[width=\textwidth,height=2.5cm]{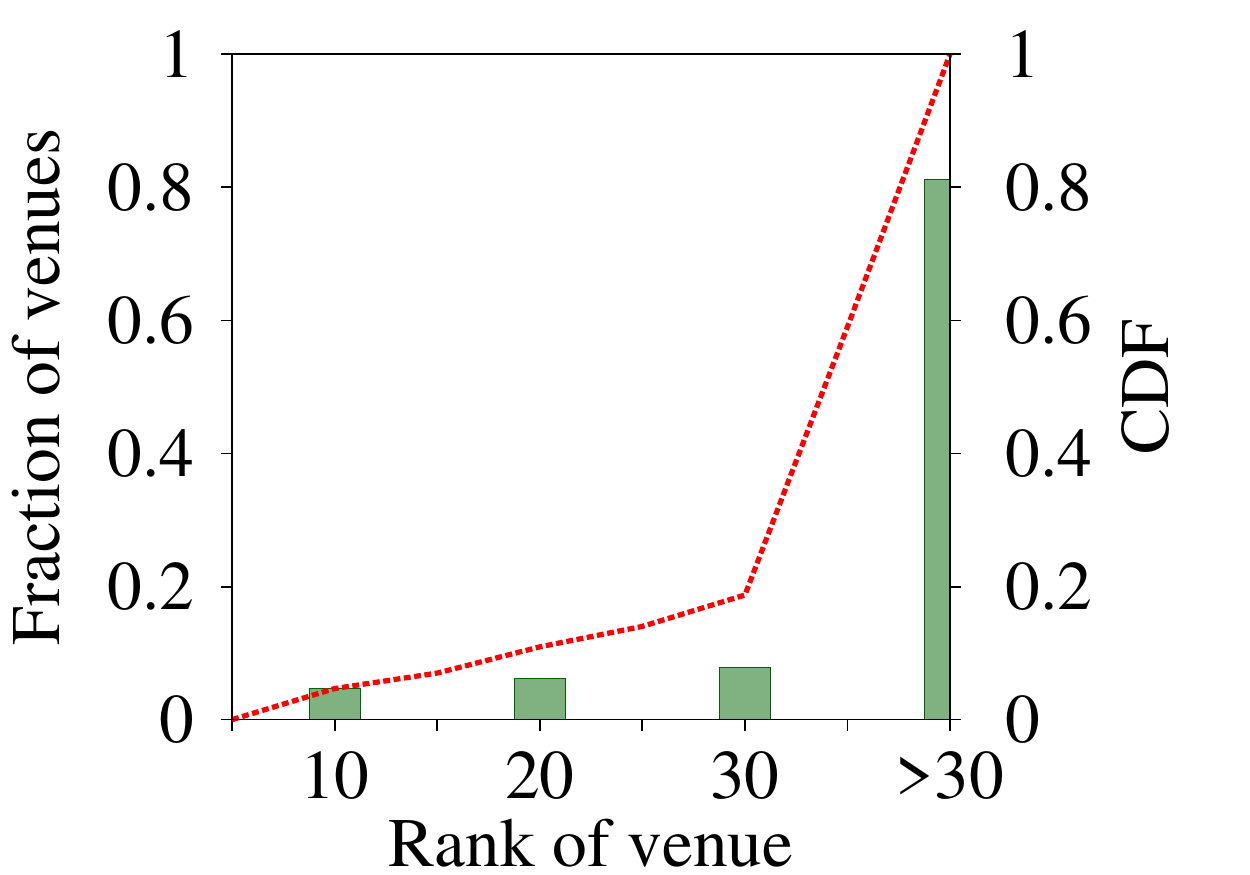}
              \caption{The rank of the actual venue in the list of nearby venues.}

        \end{subfigure}

        \caption{Quantifying the quality of the  default Foursquare ranking function (WiFi off).}
\label{fig:rankdisnowifi}
\end{figure}
\textbf{WiFi turned off:}
Fig.~\ref{fig:rankdisnowifi}(a) shows that the error in distance is larger than 200m in 45\% of cases and even worse it reaches  500m,  which happens when the top venue is outside  the actual user building.  Regarding the venues rank accuracy, Fig.~\ref{fig:rankdisnowifi}(b) shows that about 82\% of actual venues do not appear on the list of nearby venues (contains at most 30 venues) returned by Foursquare. This is due  in part to the coarse grained accuracy of cellular based localization. More specifically, we observed that  while performing check-ins at a set of neighboring venues (same block of a building), the venues lists returned from Foursquare are very similar. The most prevalent reason is that  the phone serving cell (from the cellular service provider) is the same in this block, making it difficult to identify  the user location. In addition, the venues that are always returned on the top of the list are the most popular (having the largest  number of check-ins) venues in the nearby area. Finally, about 22\%  of  the reported venues  are outdoor venues.

\subsection{Summary of Findings}

In summary, \emph{\textbf{our study highlights}} that a user will find a difficulty in finding her venue in the list and will either add a duplicate venue or not check-in at all, reducing both the system coverage and the user experience as well as missing business opportunities. This means  that there is a potential to enhance the venues rankings of LBSNs  for better user experience as well as reducing  duplicates in the LBSN database. In addition, automatic prediction of  uncovered venues names has the potential of increasing  venues coverage and reducing the granularity mismatch.

\begin{figure}
\centering
  \includegraphics[width=0.5\textwidth]{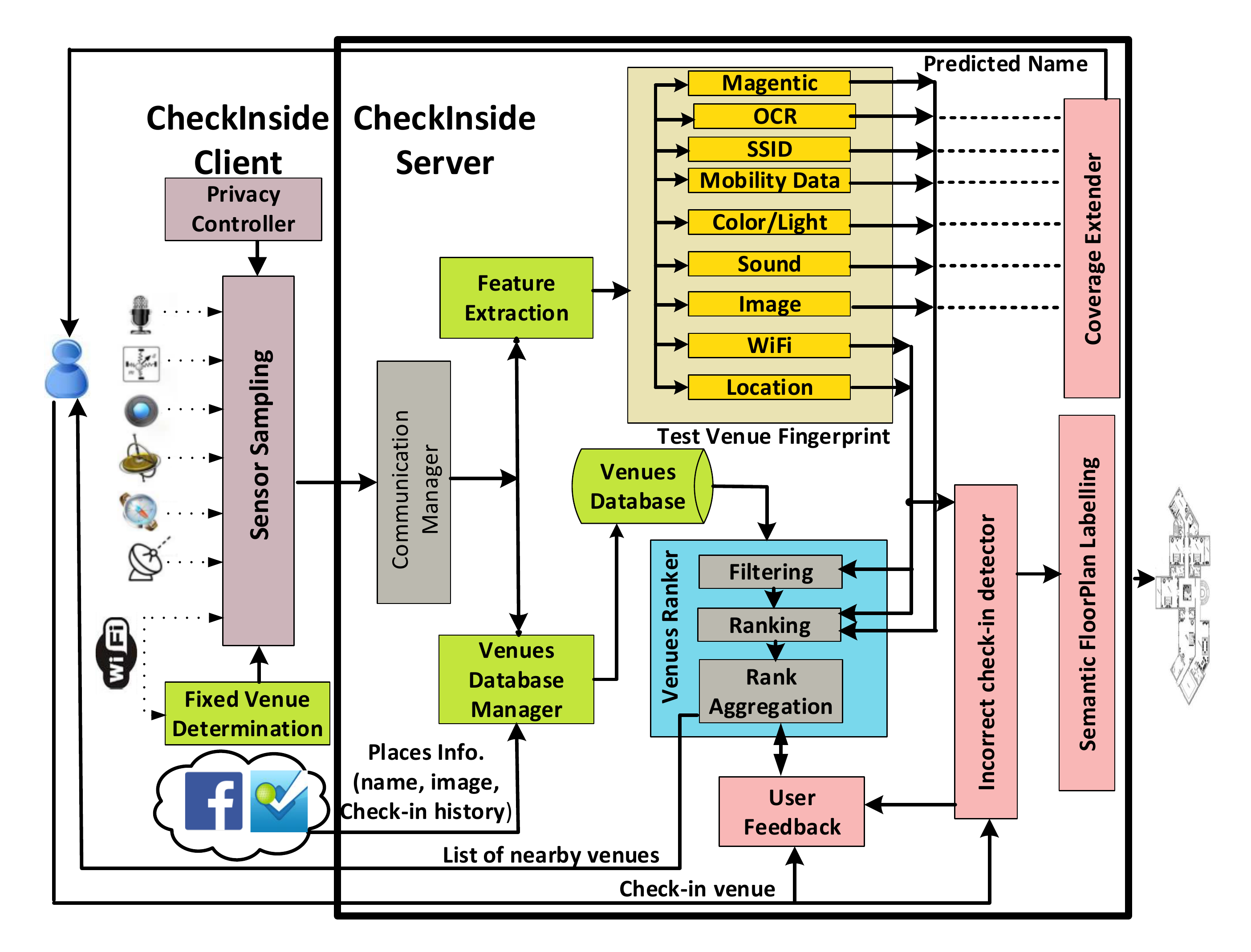}
  \caption{CheckInside system architecture.}
  \label{fig:arch}
\end{figure}

\section{System overview}
\label{sec:InCheckIn}
In this section, we present a typical scenario of how CheckInside works to illustrate the high level flow of information through the system architecture (Fig.~\ref{fig:arch}).
The CheckInside client installed on the user's phone triggers sensor data collection when a user is stable in a certain venue for some time (detected by the \textbf{Fixed Venue Determination} module). The sampled sensors  are only those  which are  enabled according to the data collection policy configured in the \textbf{User Privacy Profile}. Once a user issues a check-in request, the CheckInside client forwards the collected sensors information to the CheckInside cloud server. Sensors used are either low-energy sensors (e.g., inertial sensors), sensors that are already used for other purposes (e.g., cellular information), and/or sensors that are used opportunistically if the user turned them on for other purposes (e.g., WiFi, camera, mic). At the heart of CheckInside is an indoor localization technology. We use the Unloc system~\cite{wang2012no} due to its high accuracy, low-energy consumption, and its reliance only on the phone sensors.

Using the reported phone location, even with a coarse-grained accuracy, the \textbf{Venues Database Manager} contacts traditional LBSNs, e.g., Foursquare, to obtain a list of nearby venues and their associated information (e.g., pictures, user tips, and other check-ins data). These candidate venues are combined  with the list of nearby venues already stored in the CheckInside database and the merged list is annotated with the multi-sensor fingerprint of each venue stored in the CheckInside venues database.

The \textbf{Features Extraction Module} creates a test fingerprint of the current user location based on the collected sensors information.\\
The \textbf{Venues Ranking Module}  performs a series of accept/reject filtering operations on the returned venues  from the \emph{Venues Database Manager}  to reduce the candidate set based on the location and WiFi fingerprint by computing the pairwise similarity among fingerprints of the test  and candidate places.  It then performs a set of ranking operations, based on the different sensors employed,  to rank the candidate locations. The different rankings are then aggregated using the \textbf{Rank Aggregation Module} to produce a final ranked list of candidate locations. This list is returned to the user to select the check-in venue.

Once the user selects her current venue, the \textbf{Incorrect Check-ins Detector Module} runs to
handle outliers and noise in the user location check-in operations performed far from the actual venue. After the incorrect check-ins are removed, the \textbf{User Feedback Module} uses the correct check-ins  to update the weights of the different ranking modules to enhance the future  system performance. Concurrently, the selected user place and the test fingerprint are passed to the \textbf{Semantic Floorplan Labelling Module} to label the venue location on the map.

Finally, if the test place is not suitably matched to any  of the candidate venues in the venues database, it will be marked as a new place and  the \textbf{Coverage Extender Module} will try  to predict its name.

\section{The CheckInside System}
\label{sec:CheckInsidearc}
In this section, we present the details of CheckInside modules depicted in Fig. \ref{fig:arch}. Without loss of generality, we take Foursquare as an example of traditional LBSNs for the rest of the paper.

\subsection{Sensor Sampling Module}

This module is responsible for collecting  the sensor features from the user's phone  including the accelerometer, microphone, camera, gyroscope, magnetometer, and the received WiFi signal strength values from the available access points. The GPS is also queried with a low duty cycle to detect the user's transition from outdoors to indoors.  The   sensor measurements  are piggy-backed to the cloud server  when the user performs a check-in operation.

\subsection{Privacy Controller}

 Privacy is an important issue in the design of mobile sensing applications. People are sensitive to data captured by their phone, particularly multimedia data, and how this data is used by the application. Given this fact, CheckInside gives users full control over the sensed data by means of a personalized privacy configuration. Specifically, CheckInside has different modes of operations (full sensors, partial sensors, privacy insensitive data only, etc) that tailor the amount of data collected based on the user's preferences.
 There is a trade-off between the performance of the system and privacy. However, according to recent studies \cite{chon2013understanding},  most sensors harnessed  by  CheckInside (inertial sensors and WiFi) are enabled by most users and even  the privacy-sensitive sensors (i.e., camera and microphone) are enabled by about 78\%  of users according to the same study. Finally, we process most of the collected sensors  data locally on the user's device, further enhancing the user privacy.

\subsection{Energy Consumption}

As sensors  sampling (to capture a place fingerprint) needs several seconds, initiating it after the user starts a check-in process  will incur a high delay.  The phone sensing, therefore need to run in the background to have the place fingerprint ready when the user wants to check-in. However, continuous sensing  without duty cycling  leads to faster battery  depletion. To save  the phone battery, we apply the  adaptive sensor scheduling  scheme  \textit{triggered sensing}   \cite{mohan2008nericell}. The key idea behind triggered sensing is that sensors that are relatively inexpensive in energy consumption (e.g., accelerometer) is used to trigger the operation of more expensive sensors (e.g., camera and mic). In addition, all sensors are sampled at low rates (compass,  accelerometer and gyroscope at 24Hz; WiFi at 1Hz;  audio at 32Hz). We quantify the energy consumption of \sys{} in Sec.~\ref{sec:perf_comp}.

\subsection{Fingerprint Preparation}

This module is responsible for preparing the test fingerprint for the venue the user is currently located at as well as retrieving the fingerprints for candidate venues from the venues database. It consists of three main modules: Fixed Venue Determination, Venues Database Manager and Feature Extraction (green modules in Fig.~\ref{fig:arch}):

\subsubsection{Fixed Venue Determination}
\label{sec:venue_event}
To reduce energy consumption and enhance  users' privacy,  this module  determines  if  the user is stationary at the same venue for certain amount of time to start data collection. Since the estimated indoor location may have inherent errors that may place the user at the wrong side of a wall, i.e., another venue, we revert to using WiFi similarity for determining the stationarity within a venue, which  gives better performance~\cite{wang2012no}.
In particular, the system considers that a user is staying at the same venue if the similarity of consequently received signal strength from WiFi APs is larger than a certain threshold. We experimented with different similarity functions \cite{jaccard1912distribution,park2010growing,azizyan2009surroundsense} and found that a modified version of \cite{azizyan2009surroundsense} gives the best performance. Specifically, given two lists of APs at two locations  ($\textrm{APs}_1$) and ($\textrm{APs}_2$), the similarity is given as:
\begin{equation}
S=\frac{1}{|\textrm{APs}_u|}\sum_{a\in \textrm{APs}_u} (f_{1}(a)+f_{2}(a))\dfrac{\min(f_{1}(a),f_{2}(a))}{\max(f_{1}(a),f_{2}(a))}
\label{eq:wifi_sim}
\end{equation}
where $\textrm{APs}_u$ is the union of the MAC addresses of the APs in the two locations,
$f_{1}(a)$ and $ f_{2}(a) $ are the fraction of times each unique MAC address $a$ was
observed over all recordings in the two locations respectively. Note that this metric has the advantage of not depending on the signal strength (which varies by different devices) and, different from \cite{azizyan2009surroundsense}, is normalized to be independent of the number of APs at a particular location (it ranges from 0 and 2).

Once the user is detected to be stationary, sensors data as well as the stay duration are collected. When the user performs a check-in operation, sensor features are piggy-backed with the check-in request to the CheckInside server. Otherwise, if the user leaves the venue without performing a check-in, all venue  related data are discarded.
\subsubsection{Venues Database Manager}

This module prepares a list of the candidate venues that will be further filtered out and ranked by the Venues Ranking Module to identify the user location. It first consults the Foursquare database to retrieve the list of nearby venues given the current user location. Other data retrieved from the Foursquare database include the pictures associated with the venue, tips, check-in history, and location\footnote{The venues' location in the Foursquare database are not accurate as they are based on the outdoor GPS location or the network-based location.}. It then stores/updates this data in the CheckInside local database and retrieves the associated multi-sensor fingerprint of the retrieved list as well as the location of the venues as estimated by CheckInside, if the venue already exists in our database. It also builds an index  for brand venues (having branches in different buildings) that will be used by the Coverage Extender Module to increase  venues coverage.

Since 8\% of the venues
were registered in the Foursquare database more than once with
slightly different names (Sec.~\ref{sec:study}), to mitigate this problem for brand
venues (constituting 82\% of shops in the four malls), we
compare the list of brand names with all names registered on the venue database
based on the edit distance using the Levenshtein algorithm
for the string similarity calculation \cite{guo2014shopprofiler}. The venue name
encountering an edit distance from a brand name less than
a certain threshold is updated to the brand name.
On the other hand,
for the non-brand venues, we compare the names of all
registered venues on the same building against each other
using the edit distance. Names having low edit
distances will be clustered together as representatives of the same venue.

\subsubsection{Feature Extraction Module}
\label{sec:features}
This module extracts the features used to characterize a certain venue to generate the test fingerprint of the location the user is currently at, and is used later by the Venues Ranking Module. Features extracted cover both the user's behavior as well as surrounding environment. Specifically, we use the following features:

\textbf{\textit{Location:} }This is based on the Unloc system \cite{wang2012no} that performs dead-reckoning  to provide a rough  estimate of the phone location. To  reset the dead-reckoning accumulated error, it leverages points in the environment with unique sensors signatures (e.g., elevators, turns, etc).  Unloc has the advantages of not requiring any calibration or infrastructure, high accuracy, and low energy consumption.\\
\textbf{\textit{Mobility data:}} This group of features captures users' behavior while visiting different venues as it  is a key indicator of place category.
For example, people are stationary for a longer time in restaurants and they mostly visit them during a certain time of day (i.e., meals time). Similarly, users may go to certain shops more frequently  based on the season (e.g. ice cream shops in summer).
On the other hand, users are more mobile in clothing shops and there is no fixed pattern for the visiting time of this category. CheckInside uses three mobility features to characterize the nature of venue: (1) the user activity in the venue, (2) the timestamp (time within day)
 this type of venue is usually visited, and (3) the  time the user spends in this venue.\\
The first feature, the user activity, is define as the ratio ($r$) between the user mobility time to user stationary time within a certain period. User mobility is retrieved from  Android activity recognition service, which can distinguish  whether the user is stationary or  moving.  This is quantized into three levels: stationary (e.g., sitting in a restaurant, if $r <0.2$), browsing (e.g., in a clothing shop, if $0.2<r\le 2$), and walking (e.g., in a grocery store, if $r> 2$)\cite{lu2009soundsense}.\\
 On the other hand, visiting time is quantized into different periods within the day: early morning, late morning, early afternoon, late afternoon, early evening, and late evening\footnote{generalization to other granualities, e.g. over a week or a year, is left to future work.}.
 Finally, stay duration is  quantized into 30 minutes intervals. The fingerprint associated with the three mobility features is the histogram of the feature samples collected at this particular venue from different check-ins.\\
\textbf{\textit{ WiFi Fingerprint:}}
Due to the limited range of WiFi in indoor environments, it can be used to characterize venues indoors. CheckInside stores the fraction of times each unique MAC address was observed in the venue over all check-ins  as the WiFi fingerprint for that venue. \\
\textbf{\textit{Sound Fingerprint:}}
Sound captured by a mobile device's microphone is a rich source of information that can be used to make accurate inference about the surrounding environment. For example, some venues (e.g., a music store) play music in the background while others (e.g., a library) are quieter. To recognize venues using ambient sound, CheckInside fingerprinting is based on the signal amplitude to capture the loudness of the sound \cite{azizyan2009surroundsense}. Specifically, the amplitude is divided into 100 equal intervals and the number of samples per-interval is normalized by the total number of samples in the recording. The 100 normalized values are considered to be features of the ambient environment. Since sound from the same venue can vary over time, we divide the day into 24 1-hour bins and use a separate sound fingerprint for each bin.\\
\textbf{\textit{Image Fingerprint:}}
There are many  features used in literature to represent images including the Scale-Invariant Features Transform (SIFT) \cite{lowe2004distinctive}  and  the gist features \cite{oliva2001modeling}  which captures local  and scene features in images respectively. While these features capture the essential characteristics of images, they are not directly appropriate for our system due to their large size. For instance, each SIFT feature is a 128 dimensional vector and there are several hundred of such SIFT vectors for an image. The large size makes it inefficient for image matching, which is not suitable for the real-time operation required by CheckInside.

To resolve this problem, we leverage the visterms compact features \cite{yan2010crowdsearch} which reduce the size of the SIFT features significantly by efficient clustering. A visterm is treated as a term in a document (image in our case) which has an Inverse Document Frequency (IDF) to indicate its discriminative power. Once visterms are extracted from an image, they can be matched efficiently against visterms extracted from the images  retrieved from the venues database in a manner similar to keywords in text retrieval.

\begin{figure}
  \centering
  \includegraphics[width=\linewidth]{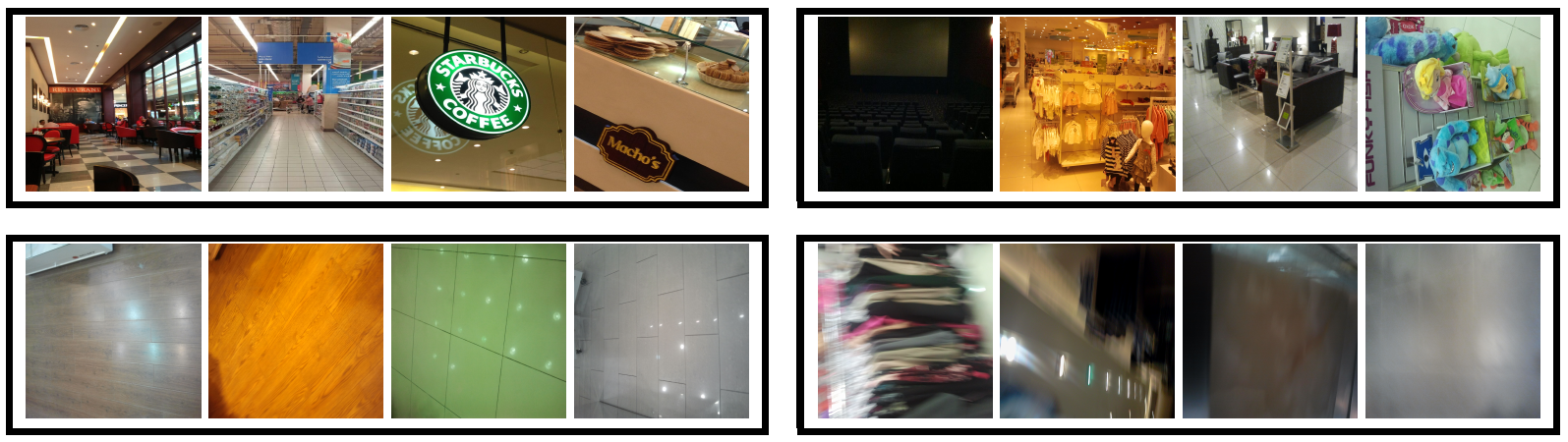}
  \caption{Pictures taken at different stores: The top left group shows example pictures used to differentiate stores (image features). The top right and bottom left groups show pictures taken at different venues with different light intensities and different floor types respectively (color/light features). Finally, the bottom right group shows some blurred images in our collection.}\label{fig:pictures}
\end{figure}

\textbf{\textit{Color/Light Fingerprint:}}
 A large number of stores have a thematic color as part of their decoration, e.g., red at McDonalds.
The wall and floor colors contribute significantly to this theme (Fig.~\ref{fig:pictures}). Floors may be covered with carpets, ceramic tiles, or wooden strips, all of which are discriminating
attributes of the ambiance.  Based on this, pictures taken from different spots in a store are likely to reflect this theme.
CheckInside extracts dominant colors and light intensity from pictures of floors and walls by transforming  the pixels of the floor images from the RGB space to the hue-saturation-lightness (HSL) space. This has the advantages of removing the effect of  shadows of objects and people, and the reflections of light; and decoupling the floor and wall colors from the ambient light intensity \cite{azizyan2009surroundsense}.

\begin{figure}
\centering
  \includegraphics[width=0.5\textwidth,height=3cm,keepaspectratio]{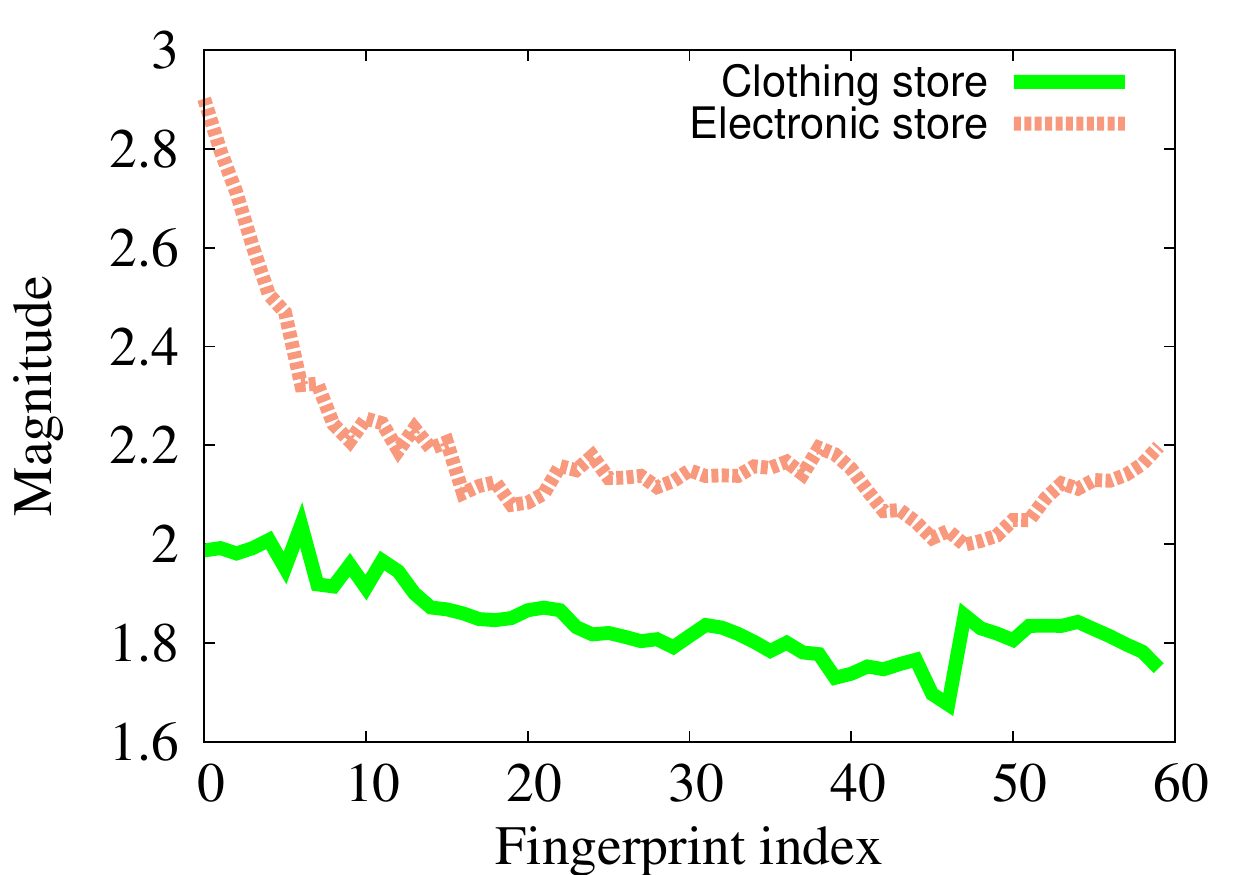} %
    \caption{Comparison between the magnitude of normalized magnetic field in an electronic and a clothing store.}
  \label{fig:magn}
\end{figure}

We run the K-means clustering algorithm on the HSL image representation of all pictures taken at the same
venue. The K-means algorithm divides the pixels into $K$ clusters, such that the sum of distances from all pixels to their centroid is minimized. The centroids of these clusters, as well
as the cluster sizes, together form the color/light fingerprint of that venue.\\
\textbf{\textit{Magnetic Fingerprint:}}
 The natural magnetic field has two  characteristics: the uniqueness of magnetic field distortion from one location to another in a building, and its  time invariance. This enables  the deployment  of magnetic-field distortion-based location estimation. The ambient geomagnetic fingerprint can be modeled as a vector M of three components $m_{x}, m_{y}$, and $m_{z}$ which represent the measured magnetic field in the  three directions x, y and z. The collected magnetic readings are normalized using mean normalization, and the magnitude of each reading is computed \cite{galvan2014magnetic}. A comparison between  the magnitude of the normalized magnetic field collected at a clothing and an  electronic venues is shown in Fig.~\ref{fig:magn}.

The FFT is applied on the normalized magnetic signature to generate an energy signature  independent of time and walking patterns. This will avoid the burden of
collecting magnetic information in different directions in order to construct an accurate magnetic map.\\
\textbf{\textit{SSID Fingerprint:}}
The SSID of an  Access  Point (AP) installed in a certain venue may be  indicative of the venue name given that the vast majority of shops have a wireless Internet connection. The WiFi scan collected while the user is performing a checks-in  contains a  number of AP SSIDs overheard at that place. Consequently, the location's SSID  fingerprint is   represented as the SSID of  the AP that has the strongest average RSS in that scan. However, before computing the strongest AP,  APs having SSID holding common  AP manufacturers  or service providers names (e.g., LinkSys or Vodafone) are filtered out.\\
\textbf{\textit{ OCR Fingerprint:}}
Opportunistic images captured while users visiting a place may contain menus, store logo, or postings. CheckInside mines words from these images  by incorporating the HP Tesseract OCR engine \cite{smith2007overview}. The set of words extracted from place-related images constitute its OCR fingerprint.  However, the venues database manager keeps track of tips posted by users at candidate venues crawled from Foursquare.\\

\textbf{\textit{Familiarity:}}
It indicates how frequently a user visits the
same venue and/or the venue's brand and  is measured by the number of check-ins
the user has performed at the venue and/or its brand. We hypothesis that
this feature will outperform popularity used in \cite{elhamshary2014checkinside}  as the
place popularity is measured over all users and a certain
venue may be popular in general but the current user is not
interested in it.
\subsection{New Venue Determination}
 As LBSNs are organic systems which are based on users' contributions, some venues may  not  be covered yet (e.g., 35\% in case of Foursquare)  as illustrated in the study shown in Sec. \ref{sec:study}. When the user issues a check-in request, CheckInside should determine if the test venue is a new (i.e., visited for the first time) one, or it is  already included in the venues database, to decide where to forward the test place fingerprint. If the test venue is new,  the  coverage extender module is used to predict its name. Otherwise, the venue ranking module is employed to match the test venue fingerprint  against candidate venues retrieved from the venue database.  To recognize whether the current  venue is new or not, we draw on the WiFi similarity (using Eq.~\ref{eq:wifi_sim}) among the current place WiFi bind and all candidate places WiFi fingerprints. If the  maximum WiFi similarity among test venue and other candidate venues is lower than a certain threshold (1.2 in our system), this indicates that the place is not covered yet by the system.

\subsection{Coverage Extender Module}
\label{sec:covext}
 Given the fact that the vast majority (82\%) of stores in large shopping malls belong to well-known international or local brands,  as shown in the study in Sec. ~\ref{sec:study}.  CheckInside leverages  this fact to extend the coverage of current LBSNs by predicting the names  of venues that are not yet  included in the LBSN database.

 To predict the name of a venue, CheckInside relies on two subsequent approaches.  The first  approach uses the WiFi scan collected  during  the check-in at the test location. It will first  determine the strongest AP at that location. The SSID of the strongest AP is compared against a predefined list of brand  names (retrieved from  the  CheckInside database). The venue name that encounters the lowest edit distance  with the strongest AP SSID  is deemed as the  predicted name for the new venue. To decrease the false positive (FP) rate, venue naming  is confirmed only if the lowest edit distance  is less than certain threshold. Nevertheless, if the first approach  failed to predict the name, the system  will compare the test place logical fingerprint (the complete fingerprint excluding WiFi, magnetic and location) against the logical fingerprints of all brand venues in other malls (indexed in the venues database).  Our intuition  is that most  shops belonging to the same brand  share decorations, color themes, lighting styles,  may play the same type of music, and have similar mobility data.  Consequently, logical fingerprints matching  of test venue against brand venues will be able to predict the correct name for brand venues with high probability.
\subsection{Venues Ranking Module}

  This component is responsible for ranking the candidate list generated by the Venues Database Manager.
 It accomplishes this  by three main components (blue in Fig. \ref{fig:arch}):
 filtering, feature-based ranking, and rank aggregation.

\subsubsection{Filtering}

The function of this component is to eliminate candidate venues that are not likely to be similar to the test venue. This helps in increasing the efficiency and accuracy of the next ranking modules. Filtering is performed based on the current user location and the WiFi fingerprint. Both filters are run independently and concurrently returning a number of candidate venues. To avoid excessive filtering,  each filter returns a fixed number of locations (taken the same as the Foursquare API default of 10\footnote{\url{https://developer.foursquare.com/docs/venues/search}}). The output lists of the two filters are aggregated, generating the candidate list. The number of places in the candidate  list  can be extended/shortened depending on the confidence in the  place inference or/and the user preference.

\textbf{\textit{Filtering By Location:}}
This is performed by placing a threshold on the distance between the current user location and the candidate venue location. The metric used for distance calculation is the shortest door-to-door walking distance, rather than the euclidean distance. To speed up the retrieval of closest candidate venue to user location, we use an R-tree to index the venues database.

\textbf{\textit{Filtering By WiFi Fingerprint:}}
It is performed by computing the similarity between the test venue WiFi fingerprint and all candidate venues WiFi fingerprints using Eq.~\ref{eq:wifi_sim} and then returning the venues with the highest scores.

\subsubsection{Feature-based Ranking}
This module orders candidate venues according to their pairwise similarity with the test venue. CheckInside employs all extracted features for  matching the test  place against candidate places.
 Each ranker orders the pruned  list of nearby venues received from the filtering component based on one of the features in parallel.

\textbf{\textit{Sound ranker:}}
To compute the degree of similarity between two  sound fingerprints, we use the Euclidean distance between the corresponding sound fingerprint 100-dimensional vectors.

\textbf{\textit{Image ranker:}}
We employ the technique developed in \cite{yan2010crowdsearch} for image search to our image ranking operation. Specifically, we use an inverted index constructed from the corpus of images in the venues database.  The inverted index is a mapping of each visterm feature in the test images to the images in the database containing that visterm. The IDF of found visterms in a candidate venue are averaged to get the venue score.
The list of candidates are returned ranked in order of their average IDF score.

\textbf{\textit{Mobility data ranker:}}
This module computes the similarity based on visiting time ($v$), user activity ($r$), and stay duration ($d$) between each venue in the candidate list and the user test venue. The similarity is taken as the joint probability of the different mobility features at the candidate venues. In particular, the mobility similarity ($m$) between the current user mobility test data ($v, r$, and $d$) and a venue fingerprint ($F$) is given by:
\begin{equation} \label{eq:mob}
m = P((v, r, d)|F)= P(F_V=v).P(F_R=r).P(F_D=d)
\end{equation}

Where $P(F_V=v)$ can be obtained from the histogram of the user visiting time at the candidate venue, $P(F_R=r)$ from the histogram of the user activity, and $P(F_D=d)$ from the histogram of stay duration.

 This metric indicates that a candidate venue is good if it has a high probability of matching the current user mobility behavior. For example, food venues would  have close visiting time (e.g., at meals time), long stay durations (e.g., 30+ minutes), and similar user activity (e.g., sitting) with high probability.\\
\textbf{\textit{Color/Light ranker:}}
The color/light similarity is performed  based on the Euclidean distance between their cluster centroids and
the clusters' sizes  \cite{azizyan2009surroundsense}. The similarity ($S$) between fingerprints $F_{1}$ and $F_{2}$ is defined as:
\begin{equation}
S=\sum_{i,j} \dfrac{1}{\delta(i,j)}  \dfrac{\textrm{sizeOf}(C_{1i})}{T_{1}}\dfrac{\textrm{sizeOf}(C_{2j})}{T_{2}}
\end{equation}
where  $ C_{1i} $, $ C_{2j} $ are set of clusters for fingerprints $F_{1}$ and $F_{2}$ respectively. $T_{1}$, $T_{2}$ are the total number of pixels in clusters in $F_{1}$ and $F_{2}$ respectively, and $\delta(i,j)$ is the centroid distance between the $i^{th}$cluster of $F_{1}$ and the $j^{th}$ cluster of $F_{2}$.\\
\textbf{\textit{Magnetic field ranker:}}
It estimates the similarity between the magnetic signal vector at a test place $M_{a}$ and a candidate place $M_{b}$ by the Euclidean distance defined as:
\begin{equation}
S(M_{a},M_{b})=\sqrt{(m_{a}^{x} -m_{b}^{x}) ^{2} +  (m_{a}^{y} -m_{b}^{y}) ^{2}+(m_{a}^{z} -m_{b}^{z}) ^{2}}
\end{equation}
\textbf{\textit{SSID ranker:}}
It computes the string similarity using the edit distance among the  SSID  of strongest AP in the WiFi bind of the test venue  and candidate venues names. For a candidate venue with duplicate names,  the average edit distance among all names of the candidate venue and the test venue name is deemed as their similarity measure.\\
\textbf{\textit{OCR ranker:}}
It  matches  venues tips against OCR text mined from images captured at the test place. As a preprocessing step,  stop words and non grammatical words are filtered out. Finally, the ranker orders candidate venues based on the number of overlapping terms between the texts extracted from OCR
and from the user tips \cite{chon2012automatically}.\\
\textbf{\textit{Familiarity ranker:}}
It  ranks more familiar venues with respect to the current user higher in the list.
\subsubsection{Rank Aggregation}

Once the different rankers provide their ranked lists of places, this module fuses them into a single ranked list. We experimented  with CombSUM  as an example of score-based methods (combine the different lists based on the assigned scores in the individual lists) and  the Borda's method as an example of order-based ones (combine the lists based on the order of places in each individual list); both representing data fusion (unsupervised rank aggregation) techniques \cite{faheem2010rank}. In addition, we experimented with different learning-to-rank (supervised rank aggregation) algorithms  like  AdaRank \cite{xu2007adarank} and Ranking SVM  \cite{joachims2002optimizing}. We compare the different techniques in Sec.~\ref{sec:perf_modules}.

\subsection{User Feedback Module}

A significant characteristic of the users' interaction with a LBSN is that the user explicitly selects a venue to check-in from the list of nearby venues, which acts as  a ``ground-truth'' for the user current place. This feedback  not only provides information about the performance of CheckInside venues ranking algorithm, but it also can improve the system future performance by identifying which ranker provides the best performance.

\noindent Specifically,  rankers performance varies from venue to another. For example, some venues have distinct color theme (e.g., restaurants), other have unique sound signature (e.g., libraries are quiet), and no WiFi signal is overheard in some venues.  After each check-in, the rankers  weight is updated to reflect the degree of user satisfaction to the returned results.
We leverage this user feedback to weigh the different rankers. Initially, to ensure correctness, fake check-ins are filtered out by the incorrect check-ins detection module. Then, we start with all rankers having an equal weight. After each check-in operation, and given that the candidate list contains $l$ venues, each ranker is assigned a score of $l-i$, where $i$ is the rank of the actual venue in the ranker's list. These scores are then normalized to add to one.

\subsection{Semantic Floorplan Labelling Module}

This component is responsible for the automatic labelling of  venue names on the floorplan. CheckInside starts with a floorplan with rooms and corridors highlighted which can be either manually uploaded or automatically generated from crowdsourced data~\cite{alzantot2012crowdinside}.
To enrich the floorplan with the semantic labels of the venue names, one \textbf{\emph{cannot simply }} use the user check-in information, which provides the current venue name and the current user location due to the errors  inherent  in the  check-in process in the form of incorrect check-ins.  So,  as a preprocessing step, the incorrect check-in detection module will identify and remove outlier and noisy check-ins taking place outside the actual venue. Once the fake check-ins are removed, the venue location is estimated as the mean of the locations of the users who check-in at this venue. Based on the law of large numbers, this mean converges to the actual location as the number of samples increases. The venue enclosing this location on the map is tagged accordingly.

\subsection{Incorrect Check-ins Detection Module}

In LBSNs, without a method for detecting fake check-ins,  the quality of location information is based solely on the honesty of users. Fake check-ins can be performed in different ways including faking GPS coordinates or checking-in a venue that is not nearby.  Moreover, the indoor localization algorithm employed has inherent error in the range of  few meters, which may place the user at an incorrect venue. Fake check-ins have a negative impact of the operations of LBSNs, including monetary loss and degraded quality (e.g., in recommendation services that make use of the check-ins data). In addition, fake check-ins affect the performance of several modules of CheckInside such as user feedback  and  semantic floorplan labelling.

To address these challenges, CheckInside uses an unsupervised outlier  detection algorithm  as there is no a-priori model available for identifying correct check-ins. Our approach is based on the outlier detection in the WiFi signal space. Specifically, we depend on the fact that independent \emph{correct} check-ins made at the same venue are adjacent in the WiFi signal space and tend to cluster, while fake check-ins are distributed over a larger area. Consequently, we apply an agglomerative hierarchical clustering approach to detect check-ins that are suspected to be erroneous. Later, label assignment  and user feedback incorporates only check-ins tagged as correct. The system maintains all  WiFi fingerprints assigned to a venue during check-ins within a time window (regardless of correctness), so that recent data can be used to periodically reclassify clusters and detect outliers for that venue. For location verification of a check-in of a user at certain venue v, we utilize the recent k RSS vectors collected by users claiming presence in v.

For the agglomerative hierarchical clustering algorithm, clusters are successively merged in a bottom-up fashion, based on the WiFi similarity metric in Eq.~\ref{eq:wifi_sim}, until the similarity falls below a pre-defined cut-off threshold $d^{*}$. The selection of appropriate value for $d^{*}$ is based on formulating the threshold identification problem as a Bayesian decision problem \cite{park2010growing}. Once check-ins are grouped into clusters, the system identifies  which cluster includes the correct check-ins (the rest are assumed to contain fake check-ins).

 If we assume that most users make correct binds, it is natural to take the largest cluster as the correct binding for the venue. However, when the system starts, it has not yet obtained enough check-ins and thus majority voting is not feasible. Therefore, we identify the correct cluster of check-ins $c_{v}^{*}$ given a set of check-in clusters ($\mathcal{C}_{v}$) at venue $v$ according to the following criterion:
\begin{equation}
c_{v}^{*}= \argmin_{c\in \mathcal{C}_{v}}\sum_{m\in \mathcal{N}(v)} d_{s}(c,c_{m}^{*})
\end{equation}
where $\mathcal{N}(v)$ is the set of neighboring venues to venue $v$, $c_{m}^{*}$ is the cluster of correct check-ins at neighboring venue $m$ at the time of computation, and $d_s(c,c_{m}^{*})$ is distance between the two clusters centroids. The intuition is that the correct cluster assignment for a venue is the one that is most similar to its neighboring venues.

\setlength{\belowcaptionskip}{-1pt}

\section{Evaluation}
\label{sec:eval}

 CheckInside is evaluated through a multi-mall deployment that include 711 stores distributed in four malls in two different cities over  six-weeks period.

\subsection{ Data Collection}

 We recruited  20 participants to collect the necessary data for evaluation. While visiting places, participants capture images and record audio samples. Simultaneously the deployed data collection tool collects user traces and samples WiFi.  To collect ground-truth (GPS is not available inside malls building),  participants manually label the venue when they depart the place. Table~\ref{tab:desc} shows the description of collected data. The light proximity sensors are used to make sure that pictures are taken when the phone is in the user's hand. The phone orientation sensor is used to differentiate between pictures of floors (used for color/light ranking) and other positions (used in images ranking). About 22.7\% of the pictures were blurred images (Fig.~\ref{fig:pictures}).

 \begin{table}[!t]
\resizebox{0.48\textwidth}{!}{
\begin{tabular}{|p{2cm}||c|c|c|c||c|c|c|}
\hline
\multirow{2}{*}{\textbf{Venue Type}} & \multicolumn{4}{|c}{\textbf{CrowdSensing}}                                                                                      & \multicolumn{3}{|c|}{\textbf{Foursquare}}                                                                                                                                                       \\ \cline{2-8}
                                     & \textbf{\begin{tabular}[c]{@{}c@{}}\# of\\ Venue\end{tabular}} & \textbf{\begin{tabular}[c]{@{}c@{}}\# of \\ Image\end{tabular}} & \textbf{\begin{tabular}[c]{@{}c@{}}\# of \\ Color\\/Light \end{tabular}}& \textbf{\begin{tabular}[c]{@{}c@{}}\# of\\ Sound\end{tabular}}  & \textbf{\begin{tabular}[c]{@{}c@{}}\# of \\ Venue\end{tabular}} & \textbf{\begin{tabular}[c]{@{}c@{}}\# of\\ Image\end{tabular}}&\textbf{\begin{tabular}[c]{@{}c@{}}\# of\\ Tip\end{tabular}}\\\hline

\textbf{Food}                  & 101        & 1366      & 1087    &380   &91     &1243 & 1299  \\ \hline
\textbf{Cloth.\&Fash.}         & 374        & 5031     & 2914     &1237  &235    &374 & 641  \\ \hline
\textbf{Arts\&Ent.}            & 23         & 287    & 183        &127    &14     &57 &138  \\ \hline
\textbf{Others}                &213         &2648      & 1514     &847    &96    & 234 &277  \\ \hline\hline
\textbf{Total}                 & 711        &9332     &5698      &2591  & 436    & 1908  &2355  \\ \hline
\end{tabular}
}
\caption{Description of collected data.}
\label{tab:desc}
\end{table}

\setlength{\belowcaptionskip}{-6pt}

Moreover, Foursquare is crawled  to extract  venues  attributes such as  name, category, user tips, among others. The dataset crawled from Foursquare contains only 436 stores (i.e., covered by Foursquare) out of the 711 stores used in this evaluation.

\subsection{Methodology}
The participants are divided into four groups. Each group of five participants is assigned to a mall. To make the data collected as natural as possible, participants were asked to behave normally while visiting each place category.  For example,  participants  browse items in the shelves and wait in queues at cashiers like normal customer at Clothing venues. Moreover, each place is visited five times on different days by different participants. The data is collected using different Android phones  including Samsung Galaxy S plus, Nexus One, Galaxy Tab, among others. This captures the time-variant nature of the fingerprint at the same venue as well as the heterogeneity of users and devices.  Finally,  some participants check-in  in the neighborhood of the place (e.g., corridors) while others check-in inside the place.

\subsection{Performance Results}

We start by evaluating the accuracy of different system modules and the performance of the system in different modes of operation. Finally, we quantify the advantage of CheckInside as compared to traditional LBSNs, in terms of ranking accuracy, coverage and power consumption.

\subsubsection{Performance of Different System Modules}\label{sec:perf_modules}

\begin{figure}[!t]
\noindent\begin{minipage}[t]{0.485\linewidth}
  \includegraphics[width=1\textwidth,height=2.8cm]{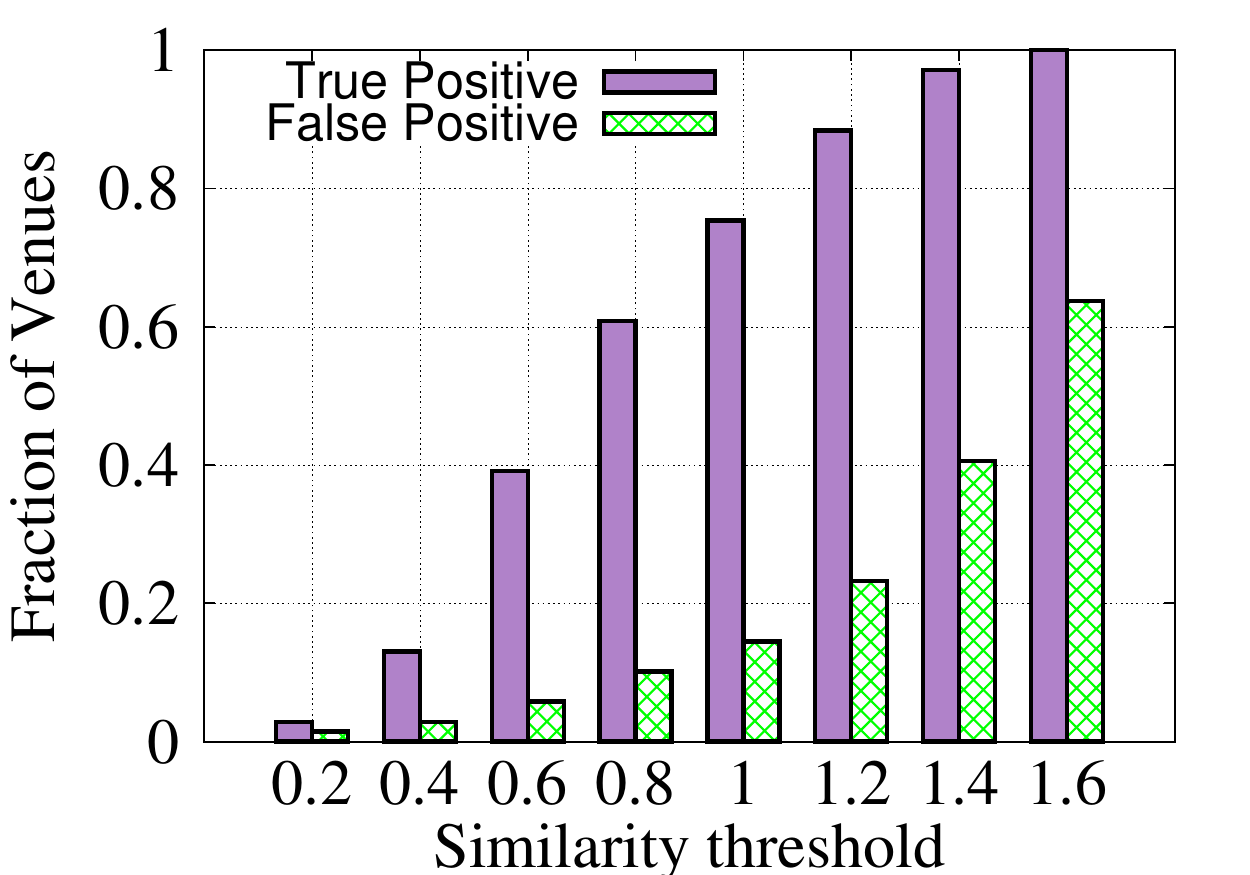}
  \captionof{figure}{Performance of  new venue recognition.}
  \label{fig:newvenue}
\end{minipage}
\hfill
\begin{minipage}[t]{0.485\linewidth}
\includegraphics[width=1\textwidth,height=2.8cm]{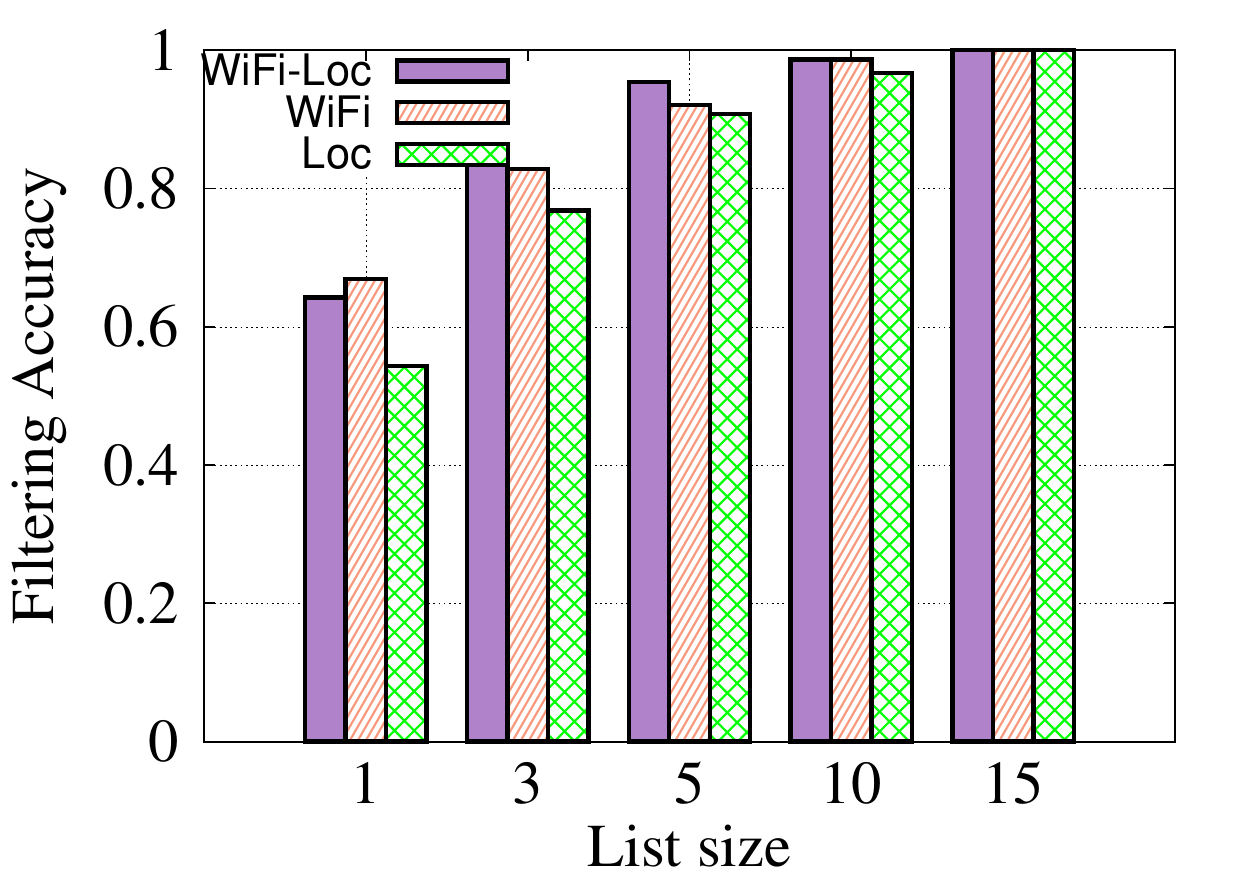}
\captionof{figure}{Performance of the different venues filters.}
\label{fig:filter}
\end{minipage}
\end{figure}

\noindent\textbf {1) New Venue Determination Module:} To evaluate the ability of CheckInside to recognize new places  (not included in its database), we perform  a  two round process. At the  first round, a WiFi fingerprint of each  venue is  matched against all WiFi fingerprints of other venues in the database (not including itself)  simulating the case when the user is at a new place. At the second round,  we match the WiFi fingerprint of each  venue against all  other venues in the database (including itself) simulating the case when the venue is covered by the system. At each case, the \textbf{maximum WiFi similarity} is calculated and compared against a predetermined similarity threshold. If the maximum similarity is less than the similarity threshold, the venue is confirmed as a new venue. Fig.~\ref{fig:newvenue} shows the effect of varying the similarity threshold (from conservative to lenient) on the CheckInside ability to recognize new venues. When the system is conservative (i.e, small threshold) in recognizing new places, it can avoid  False Positive (visited places classified as new one).  As the threshold increased, the  module will  recognize more new venues but  at the cost of higher FP rates. A similarity  threshold of 1.2 is selected  empirically  as a comprise between False Positive (FP) and True Positive (TP) rates.

\noindent\textbf{2) Venues Ranking Module:} \\
\textbf{Filtering:} Filter accuracy refers to the ability of the filter to return the actual venue within its list. Fig.~\ref{fig:filter} shows the effect of changing the candidate venues list size on the filter accuracy. It shows that both the WiFi and location-based filters have comparable performance with a slight advantage to WiFi-based filtering due to the wall-aliasing effect described before. The accuracy can be further increased by combining their output. CheckInside can achieve 100\% accuracy with a candidate list as small as 15 entries. This is compared to Foursquare that can achieve only 29.5\% for the same list size as we presented in our earlier study. This highlights that CheckInside can enhance the user experience, especially for mobile devices with small screens. This is \textbf{further} enhanced by the ranking modules.

\begin{figure}[!t]
\noindent\begin{minipage}[t]{0.485\linewidth}
  \includegraphics[width=1\textwidth,height=3cm]{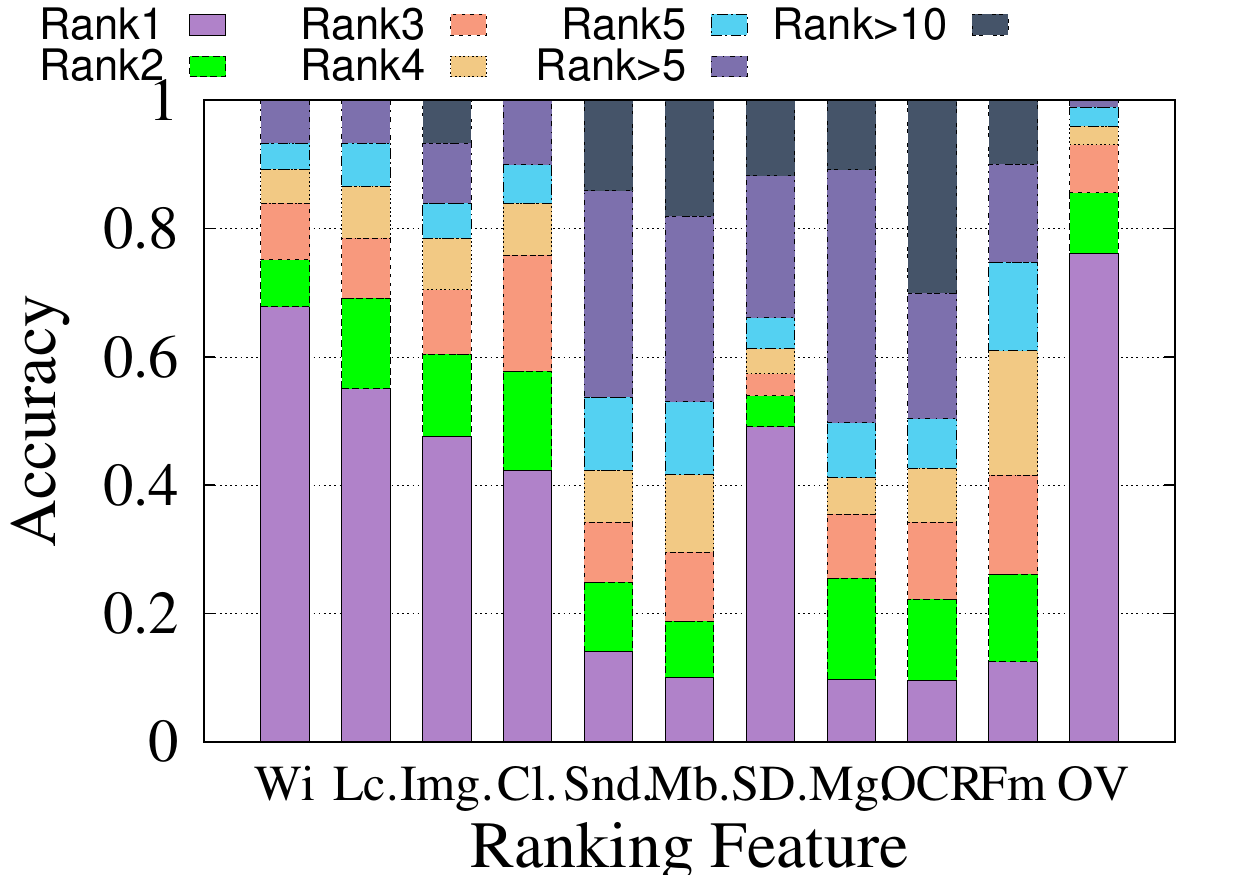}
  \captionof{figure}{Performance of the different rankers.}
  \label{fig:rankers}
\end{minipage}
\hfill
\begin{minipage}[t]{0.485\linewidth}
\includegraphics[width=1\textwidth,height=3cm]{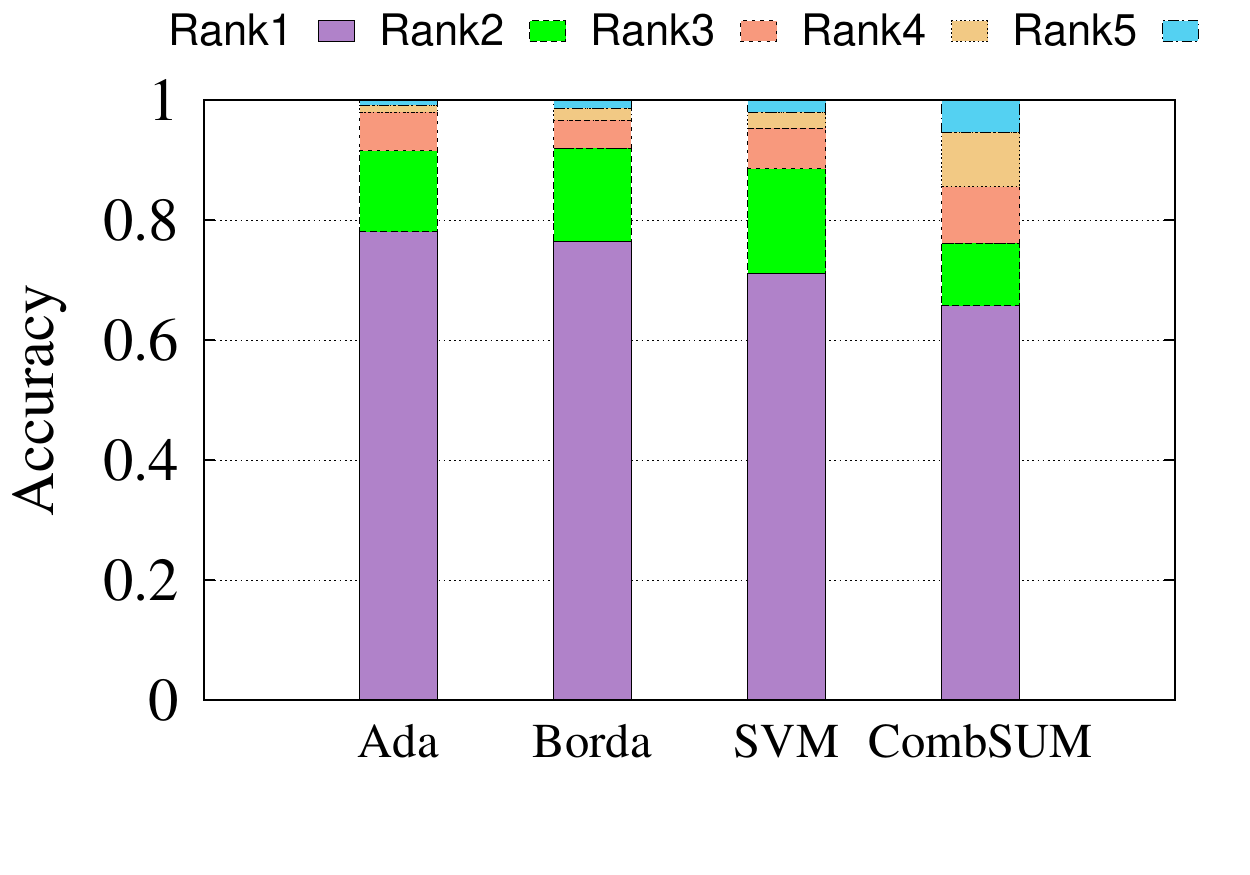}
\captionof{figure}{Comparison of rank  aggregation techniques.}
\label{fig:agg}
\end{minipage}
\end{figure}

\textbf{Ranking:}
The performance of different rankers in isolation  is shown in Fig.~\ref{fig:rankers}.

The WiFi and location-based rankers have the best accuracy due to relatively lower noise compared to the other rankers. The WiFi ranker, again, has better performance than the location ranker due to the wall-aliasing effect.

  For the Color/Light  ranker, the average accuracy for detecting the user venue is about 42\%. We observe that this ranker performs well in stores where there is a large diversity in the color and light intensities. For example, food places  are  usually grouped  into food court area and thus for economical competition they have diverse color themes and decorations.\\
 The image ranker achieves moderate accuracy (47.6\%). This is due in part to the 22.7\% collected blurred images. Another reason that made the image fingerprint less effective is that most of adjacent venues are of the same category and the majority of venues (over 50\%) are clothing and fashion shops. Therefore, pictures captured in one shop are similar to other  pictures taken at adjacent shops as they contain similar shelves and items. We noticed that the image fingerprint achieves high accuracy in small size shops, where the majority of the captured pictures have a number of common interest points due to the limited area available  to take photos.\\
The sound ranker offers a  little discriminative power as the majority of stores have a small size at which the sound samples recorded at a venue contain noise from adjacent stores. Moreover, the majority of shops are crowded with people, which makes the sound fingerprint has similar background noise for different stores. Finally, we observe  that the majority of stores are correctly identified based on their sound fingerprint are those that either have no music played in the background or are less popular venues having a small number of customers.

For the mobility ranker, the logical design of the malls, which divides the available space into blocks mostly from the same category prevents the diversity of mobility data (user activity, visiting time, and stay duration). This leads to ambiguity between adjacent stores. Another reason is that the stay duration is highly affected by the time at which  users perform the check-in. For example, if a user performs a check-in just after arriving a cafe (typically identified by long stay periods), this will make the reported stay duration short, leading to incorrect identification.\\
The magnetic ranker  provides  the actual venue on the top-5 of the list in 49.5\% of cases. This degraded performance is due to the difference of the movement pattern of different users which affect magnetic field distortion sensed by the phone compass. The magnetic field  can perform better when the magnetic field is measured  at each point in all direction forming a complete circle as discussed in \cite{chung2011indoor} which is impractical. However, it can easily recognize  stores  having a distinct pattern of magnetic distortion by metals  and/or electronics (e.g., electronic stores).\\
The OCR based ranker orders the actual venue on the top-10 of the list in 74\% of cases.
Its  performance  depends on the amount of words mined from images, the amount of tips as well as their overlapping  terms. Consequently, it achieves best accuracy in cafe and restaurant   which contain large number of tips conveying user feedback about   different menu  items and the images uploaded or collected opportunistically at these venues type always contain  menus and/or store signage.\\
The SSID ranker performs well  and it can correctly infer the user location in about 49\% of cases.  The reason is that  58.7\% of venues have  APs conveying SSIDs  very similar to venues name.  However, about 18\% of these AP are not the strongest SSID in this venue.

Finally, familiarity  ranking usually used in Foursquare does not achieve the desired performance. This due to the unpredictable  nature of user behavior and  the fact that on average 50\% of generated  of a venue check-ins  are  performed by  first time visitors (Sec.~\ref{sec:study}). The familiarity ranker, however, may be important as a tie-breaker when other features are absent or they reported approximately the same similarity values.

Finally, the last bar in Fig.~\ref{fig:rankers} shows the CheckInside rank aggregation performance with \textbf{equal} weights for all rankers. The figure shows that the actual venue is within the first five places 99\%  of the time. This is even enhanced using the feedback module as quantified next.\\
\textbf{Rank Aggregation:} Fig. \ref{fig:agg} compares the performance
of different rank aggregation algorithms. Evident from the
figure, for data fusion algorithms, the Borda's method
provides better performance than CombSUM as rankers
have high score variance which may make one ranker
dominate the others, even if it is not the best ranker.
For learning-to-rank algorithms, AdaBoost outperforms
Ranking SVM as it is designed to optimize loss functions
incorporated with any performance measure.  Overall, AdaRank
outperforms all rank aggregation algorithm with slight
improvement over Borda's method. However, we opt to use
Borda's method as it does not need training, is simple
to implement, and does not require parameters tuning.

\begin{figure}[!t]
\noindent\begin{minipage}[t]{0.485\linewidth}
  \includegraphics[width=\textwidth,height=2.8cm]{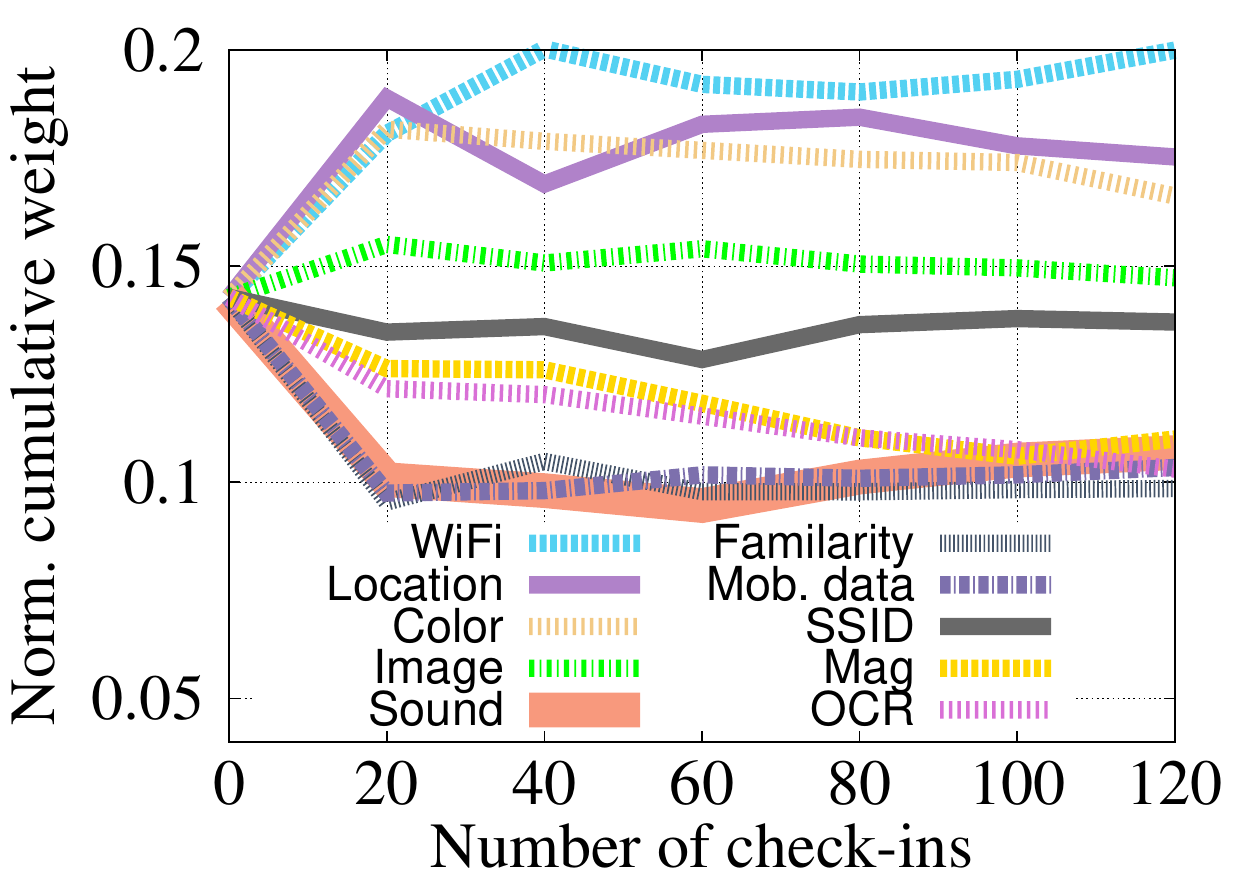}
  \captionof{figure}{Weight evolution of different rankers starting from equal weights.}\label{fig:equal}
\end{minipage}
\hfill
\begin{minipage}[t]{0.485\linewidth}
\includegraphics[width=1\textwidth,height=2.8cm]{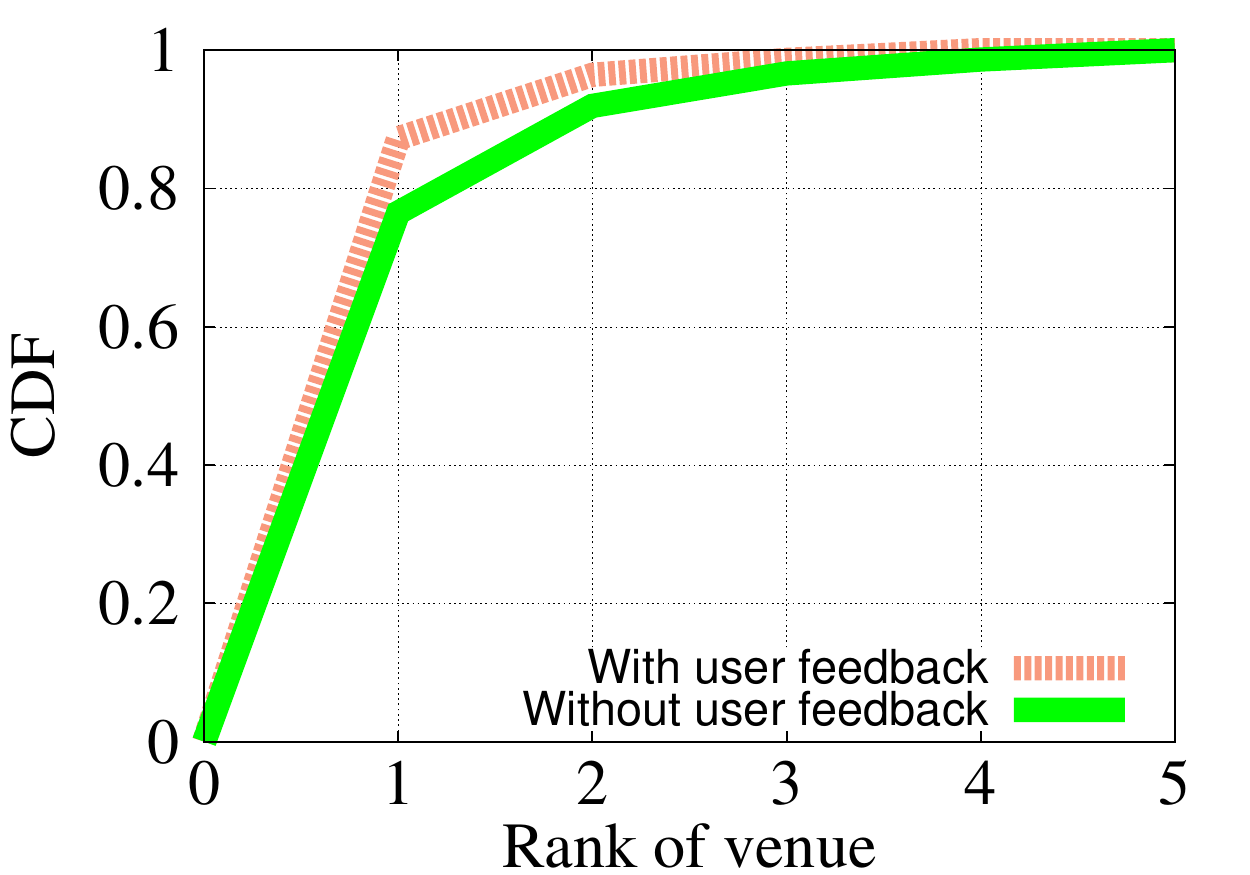}
\captionof{figure}{The CDF of the rank of the actual venue in final list.}
\label{fig:feed}
\end{minipage}
\end{figure}

\noindent\textbf{3) User Feedback Module:} Fig.~\ref{fig:equal} shows how the weights of the different modules evolve with time starting from equal weights. The weights converge to values proportional to the discriminative abilities of the different rankers as quantified in the previous section. This is reflected on the overall accuracy as shown in Fig.~\ref{fig:feed}, where using the feedback module can  detect the actual venue 87\% of the time  compared to only 76\% without using user feedback.

\begin{figure}[!t]
\noindent\begin{minipage}[t]{0.485\linewidth}
  \includegraphics[width=1\textwidth,height=2.8cm]{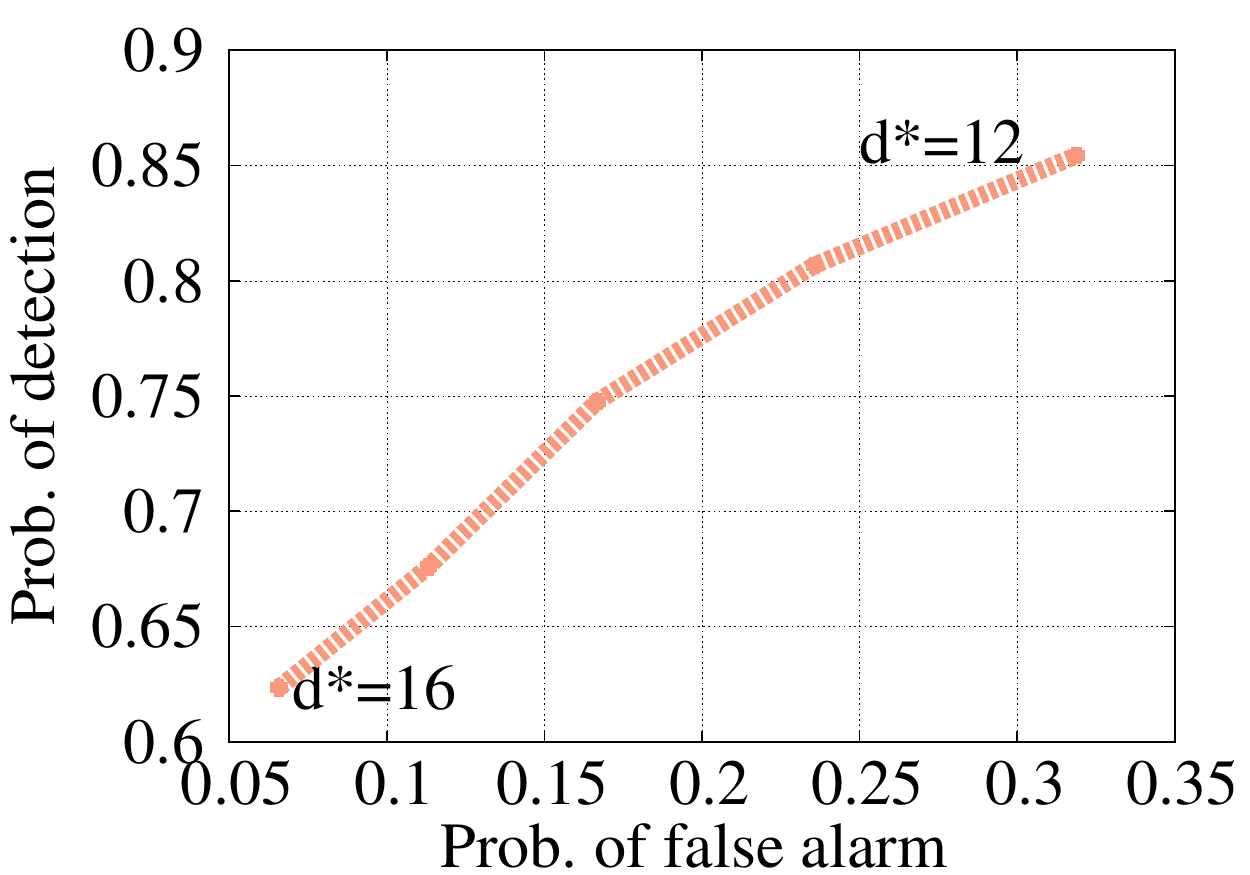}
  \captionof{figure}{The performance of fake check-in detection at different  clustering thresholds.}\label{fig:fake}
\end{minipage}
\hfill
\begin{minipage}[t]{0.485\linewidth}
\includegraphics[width=1\textwidth,height=2.8cm]{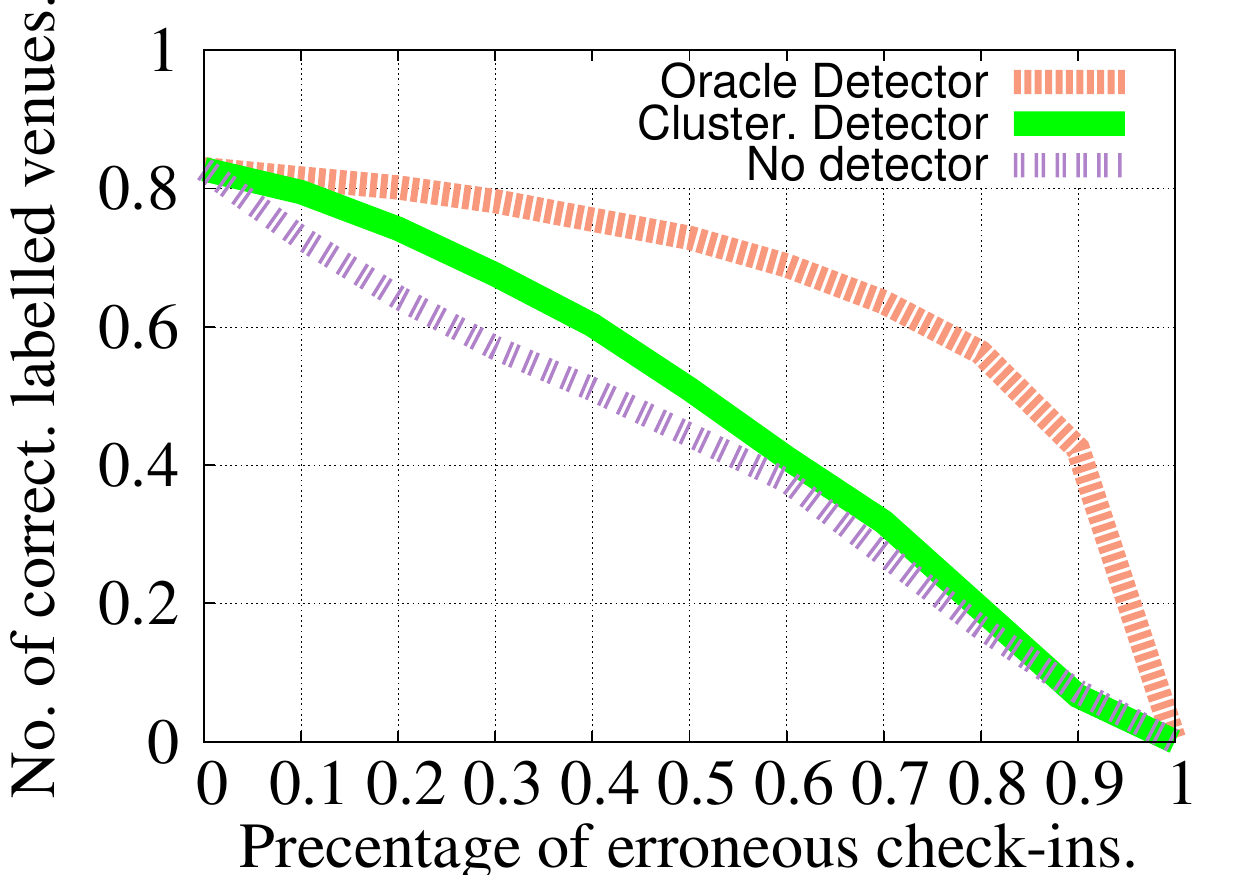}
\captionof{figure}{Semantic labelling accuracy at different check-in errors ($p_e$) for different detectors.}
\label{fig:semantic}
\end{minipage}
\end{figure}

\noindent\textbf{4) Incorrect Check-in Detection Module:} Fig.~\ref{fig:fake} evaluates the performance of the fake check-in detection  at different clustering threshold values $d^{*}$  ranging from conservative ($d^{*}$=12dB) to lenient ($d^{*}$=16dB) clustering. It shows that the trade-off  between  the probability of correct outlier detection and false alarm (a correct check-in classified as an outlier). Larger value of $ d^{*} $ means that a WiFi bind  has high probability to join clusters which decreases the probability of detection. This is due to some fake check-ins will fall into true clusters. However, using large threshold value decrease the probability of false alarm.

\noindent\textbf{5) Semantic Floorplan Labelling Module:} To understand the accuracy of the outlier detection technique of CheckInside regarding correctly inferring the venue location on the floorplan, we select for each venue a group of erroneous check-ins (performed at corridors or at other shops), where $p_{e}$ represents the probability of erroneous check-ins. We compare to an ``Oracle'' (perfect) detector.
Fig.~\ref{fig:semantic} shows that the outlier detection algorithm of CheckInside can provide from 9 to 19\% enhancement in detecting the actual venue location over a wide range of $p_e$. Note that, even with the Oracle detector, there are still errors in labelling the venue due to the inherent errors in the location determination system used. Unsurprisingly, at high error rates, the detector has a low accuracy as all detected neighbors have errors. We believe, however, that the performance is reasonable to typical values of $p_e$.

\begin{figure}[]

        \centering
        \begin{subfigure}[t]{0.24\textwidth}
                \includegraphics[width=\textwidth,height=2.8cm]{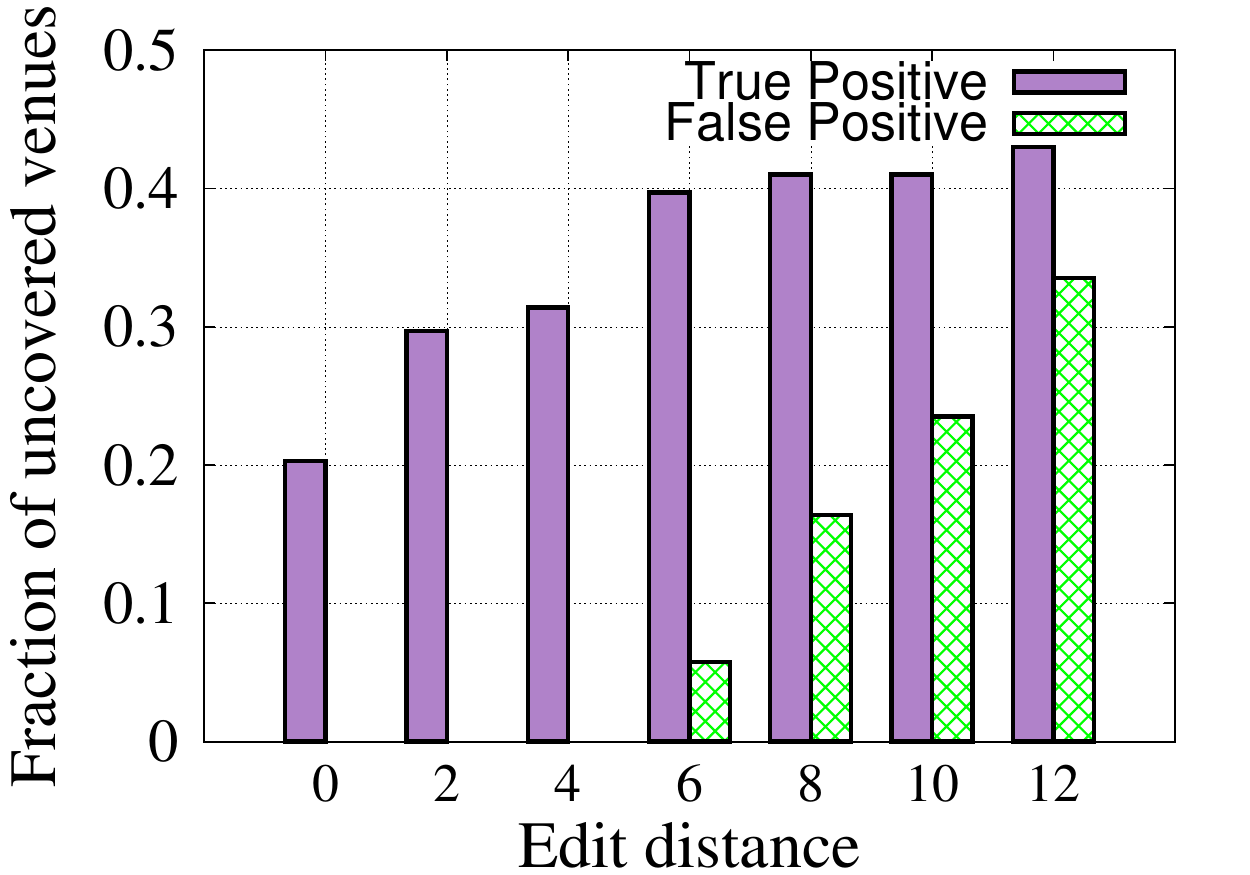}
        \caption{The prediction accuracy of uncovered venues names at different edit distance thresholds.}
        \end{subfigure}
        \begin{subfigure}[t]{0.24\textwidth}
                \includegraphics[width=\textwidth,height=2.8cm]{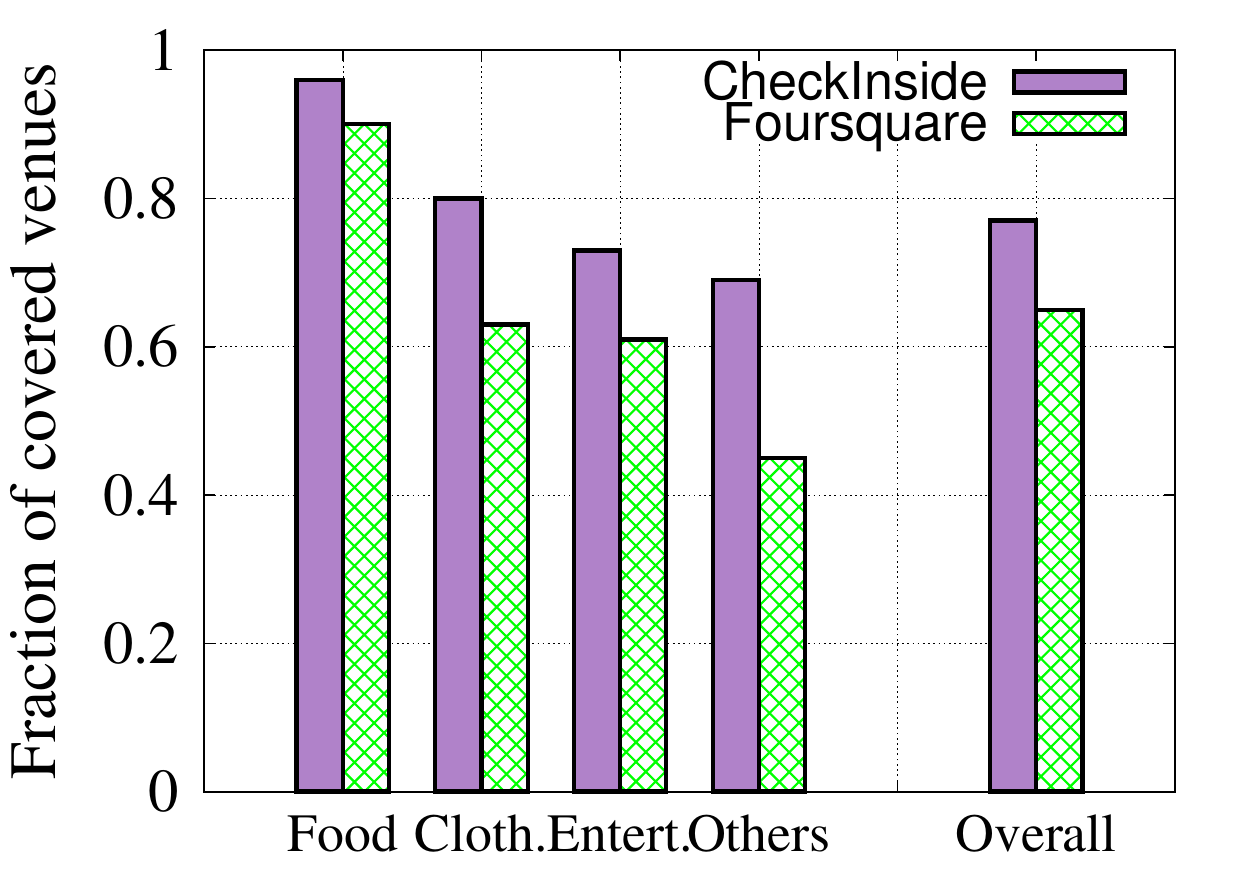}

        \caption{Extended coverage using CheckInside for different categories compared to Foursquare.}
        \end{subfigure}

        \caption{Evaluation of the coverage extender module.}
\label{fig:cov}
\end{figure}

\noindent\textbf{6) Coverage Extender Module:}
The coverage enhancement is evaluated as  the percentage of venues that are discovered by CheckInside and are not  included in the Foursquare database. To evaluate the ability of CheckInside to increase the coverage of current LBSNs, we performed a cross validation experiment where all the venues of one mall are used as the test and the venues of the other three malls are used as the database.
 The ability of CheckInside to predict venues \textbf{names}  using APs SSIDs  is shown in Fig.~\ref{fig:cov}(a). The Fig. shows the  prediction accuracy at different values of edit distances  among venues names and strongest AP's SSID. As the edit distance  increased, more venues name can be predicted at the cost of false prediction. When the distance threshold reaches a certain value, no more venues are added as the coverage extension is bounded by the number of brand  uncovered venues. Eventually, the system is able to predict the names of  about 31\% of uncovered venue without any false prediction. Fig.~\ref{fig:cov}(b)  shows the logical localization based name prediction accuracy by plotting venues coverage ratio  for different venues categories. It  shows that this approach  can  predict the names of 34\% of uncovered venues increasing  the coverage ratio of the system from 65\% (the Foursquare ratio including the granularity mismatch) to 77\%. The logical localization based prediction accuracy depends on the category and it is better for categories that are more likely to have many brand venues and has a diversity in their ambient fingerprint, e.g., restaurants chains.

  Finally, using the two approaches subsequently  will allow CheckInside  to  extend the coverage of traditional LBSNs by 37\%.

\subsubsection{Performance in Different Modes of Operation}

Fig.~\ref{fig:modes} shows the performance of CheckInside in different modes of operation that can trade-off accuracy and privacy. In particular, we compare the full system using all sensors (Full) with using only location information derived from the indoor localization technique (Loc. only), and the system when the camera is turned off (Cam. off), mic is turned off (Mic. off) and when both are tuned off (Mic.-Cam. off). The figure shows that CheckInside can provide 58\% better performance in estimating the exact venue than a simple system that is based only on location estimation. This confirms the usefulness of the semantic fingerprint. The figure also shows that CheckInside can maintain high accuracy in detecting the exact venue location (71\%) even when the privacy-sensitive sensors are turned off.

\subsubsection{Comparison with other Systems } \label{sec:perf_comp}
We compare the performance of CheckInside to Foursquare (as a typical LBSN) and  the place naming approach in \cite{chon2013autonomous} that extends traditional LBSNs to do semantic fingerprint matching using the information available about the venues from LBSN (i.e.,  visit time, popularity, tips, and images SIFT and gist features) but without taking the venue physical location nor the phone sensors (except the camera) into account for places inference.

\begin{figure}[!t]
        \centering
        \begin{subfigure}[b]{0.24\textwidth}
                \includegraphics[width=\textwidth,height=2.8cm]{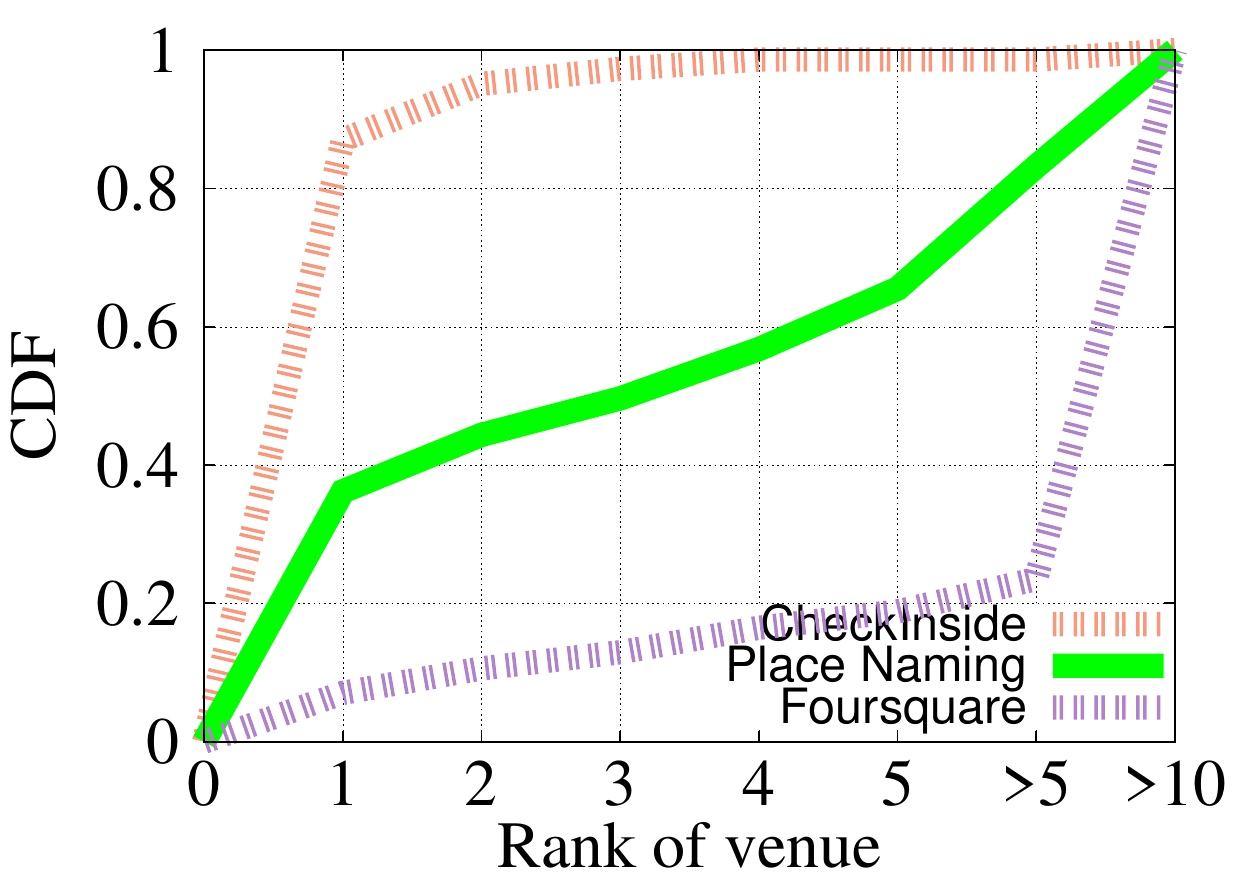}
            \caption{Rank of actual venue.}

        \end{subfigure}
        \begin{subfigure}[b]{0.24\textwidth}
                \includegraphics[width=\textwidth,height=2.8cm]{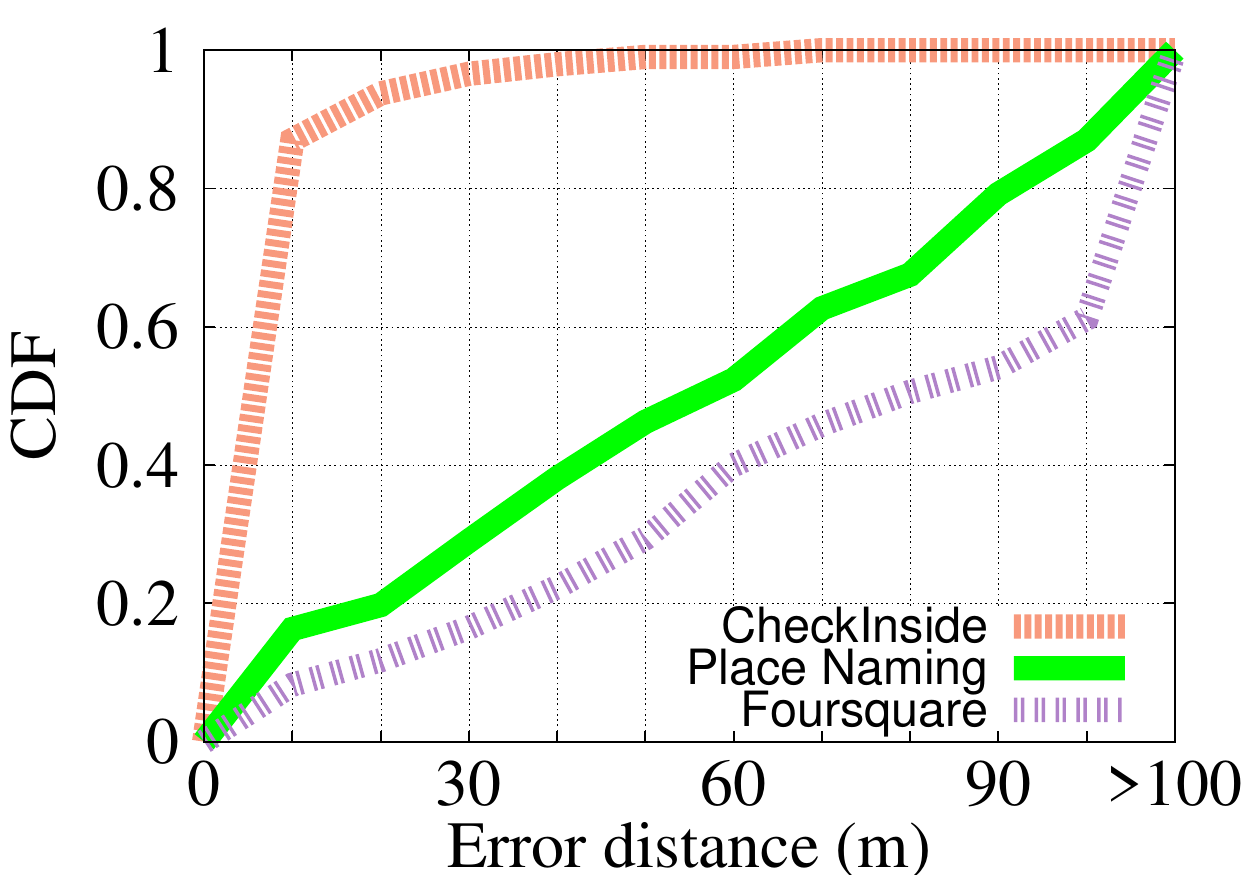}
               \caption{Distance error.}

        \end{subfigure}

        \caption{Comparing ranking of CheckInside against other systems.}
\label{fig:total_comp}
\end{figure}

As illustrated in Fig.~\ref{fig:total_comp}(a), CheckInside correctly inferred the exact venue in 87\% of the cases compared to less than 36\% for the best of the other two systems. Moreover, CheckInside can infer the correct venue within the top 5 venues 99\% of the time as compared to 62\% for the closest system, leading to a better user experience.

Fig.~\ref {fig:total_comp}(b) shows the distance error between the actual venue and the top venue suggested by different systems. As evident from the figure, CheckInside median distance error is less than 7m as compared to less than 52m for the closest systems. This highlights that CheckInside can provide better value for location-based business.

\begin{figure}[!t]
\noindent\begin{minipage}[t]{0.48\linewidth}
  \includegraphics[width=1\textwidth,height=2.6cm]{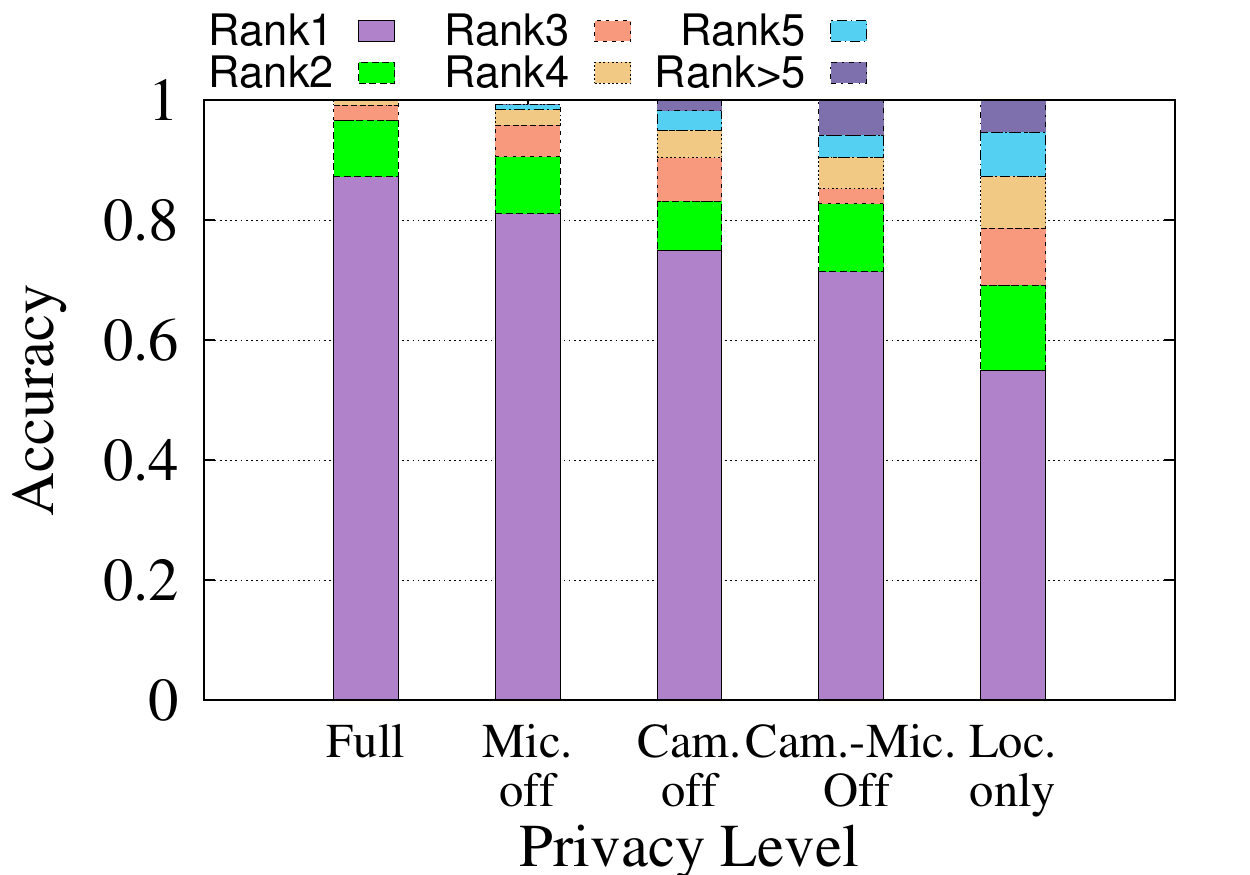}
 \caption{The  actual venue ranking for different modes of operations of CheckInside.}
\label{fig:modes}
\end{minipage}
\hfill
\begin{minipage}[t]{0.48\linewidth}
\includegraphics[width=1\textwidth,height=2.6cm]{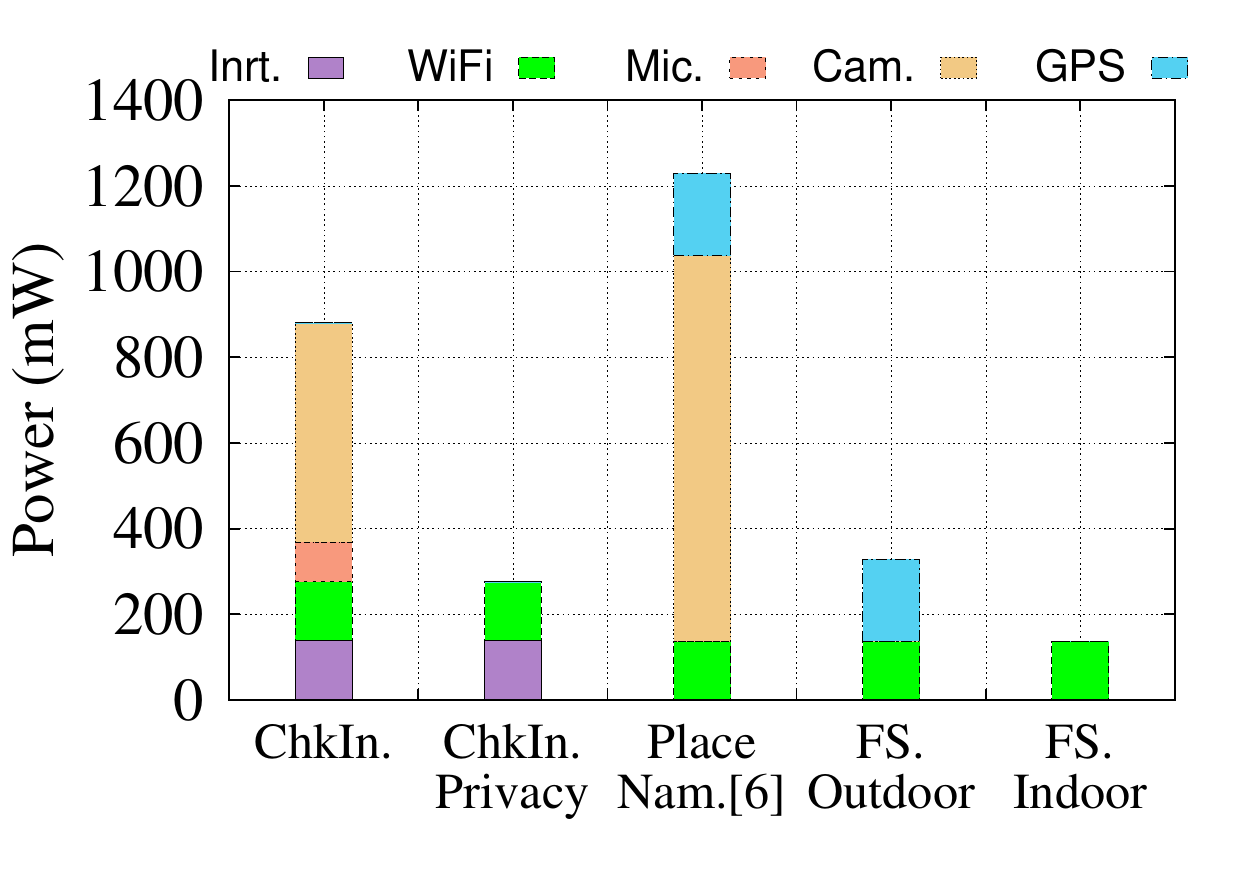}
\caption{The energy footprint of \sys{} and other systems.}
\label{fig:ener}
\end{minipage}
\end{figure}

Fig. \ref{fig:ener} compares the power consumption of CheckInside with the other two systems. The
power is calculated using the PowerTutor profiler \cite{gordon2013powertutor}  and
the Android APIs using the HTC Nexus One cell phone.
The figure shows that the camera is the most energy hungry sensors. However, due to the \sys{} triggered sensing scheme, its energy consumption is still better than other techniques that use the camera continuously.
Moreover, when the camera is turned off, e.g. in the privacy-preserving mode, \sys{} power consumption is comparable to the most power efficient system, with superior accuracy advantage (Fig.~\ref{fig:modes}). In addition, noting that the inertial sensors are always running to detect the phone orientation changes and WiFi may be already turned on by the phone user, \sys{} typically consumes little extra energy in addition to the normal phone operation.

\section{Related Work}
\label{sec:related}

Previous work focus on three main categories: venues discovery, venues description, and venues recommendation.

\textbf{Venues Discovery}
means to learn significant locations that are semantically meaningful to people such as home or work. Techniques
  \cite{kim2011towards,zheng2010collaborative} mine the user GPS trajectory and other extracted features (e.g., duration spent at each location) to infer this information. Other techniques, e.g., PlaceSense \cite{kim2009discovering},  use RF beacons to extract the visited places, while others, e.g., SenseLoc \cite{kim2010sensloc}, rely on both WiFi and GPS; and accelerometer to detect visited places and distance travelled by users. CheckInside complements these techniques by providing finer-grained labelling and categorization of indoor locations with richer semantics.

Another category of techniques, e.g., SurroundSense\cite{azizyan2009surroundsense}, use  \textbf{manually-created} semantic fingerprints to infer the user \textbf{location}. CheckInside, on the other hand, automatically creates these fingerprints in a way transparent to the user (i.e., more ubiquitous). Moreover, it addresses the incorrect association between the current user location and the check-in venue. Finally, it provides a complete system for both indoor LBSNs and semantic floorplan labelling.

\textbf{Venues Description}
refers to assigning either category (e.g., restaurant, drug store)\cite{chon2012automatically,semsense}, business naming (e.g., KFC)\cite{chon2013autonomous}, informal labels (e.g., my hometown) \cite{pontes2012we}, or activities or function associated with the location (e.g., eating, playing football, PoIs) \cite{lian2011collaborative,transit,map++,dejavu,semanticslam} to venues.
Some of venues description techniques  depend on data collected from smartphones by  crowd-sensing, e.g., \cite{chon2012automatically,lian2011collaborative}. The CSP system \cite{chon2012automatically} exploits opportunistically captured images
and audio clips from smartphones to link place visits
with place categories (e.g., store, restaurant). Other techniques, e.g.,
 \cite{pontes2012we}, exploit existing large-scale data collections (e.g., Yelp PoI database) or location-based community-generated content (e.g., Foursquare). The closest work to ours is the place naming technique in \cite{chon2013autonomous} that integrates OCR text from images, mobility data, and image features with  information about venues available from social networks to provide category (e.g., food), business (e.g., KFC), or personal naming (i.e, home and work) of venues visited by users. This system, however, focuses only on the \textbf{offline analysis} of data collected within known venues (i.e.,\textbf{ it assumes zero localization error}) and does not distinguish between indoor an outdoors venues. CheckInside, on the other hand, targets indoor LBSNs, exploits more sensing dimensions available in  mobile devices to address intentional or accidental location and check-in inaccuracies and enhances the ranking performance of LBSNs, is oriented for indoor operation, and works in real-time to provide the user with an accurate list of nearby venues. Moreover, CheckInside generates a semantic floorplan of  indoor buildings and leverages the user implicit feedback in the check-in operation to adapt the  system dynamically. Finally, our system increases the venues coverage percentage of current LBSNs.

\textbf{Venues Recommendation}
With millions of users of LBSNs, rich knowledge has accumulated about places visited by users  which enables  suggesting meaningful locations to the user. In \cite{park2007location}, the system matches users
profile data (age, gender, cuisine preferences, and income) against the price and
category of a restaurant using a Bayesian network model. Other systems, e.g., \cite{ye2010location}, exploit users' rating for places available online (e.g.,  Yelp) to suggest places to their user; while other recommendation system \cite{cao2010mining} employ user-generated trajectories to find interesting locations.
CheckInside can provide richer and fine-grained venue information, which can be used by these system to provide better recommendations.

\section{Discussion }
\label{sec:disc}

\textbf{Other Applications:}
CheckInside is currently applied to shopping malls. However, the same concept can be extended to other venue types (e.g., educational buildings). In addition, the user visiting patterns can be analyzed to provide valuable statistics (i.e., indoor analytics).

\noindent \textbf{Energy-Performance Tradeoff:} CheckInside fuses different
phone sensors to identify the user current place accurately,
which may consume more energy. To further reduce
energy consumption, in addition to triggered sensing and leveraging energy-efficient sensors, users can select energy-friendly
modes (e.g. disable energy-hungry sensors like camera and
the sound sensor), possibly reducing the system accuracy.
Nonetheless, CheckInside still outperforms the
of other systems in this case with significantly better energy consumption (Figs. \ref{fig:modes} and \ref{fig:ener}).

\section{Conclusion}
\label{sec:conclude}
We presented the CheckInside  as a  fine-grained indoor location-based social network. CheckInside leverages  data mined opportunistically from the users' phone sensors and data about venues  extracted from traditional LBSNs to fingerprint each venue. It then applies a number of filtering and ranking steps to create a ranked list of candidate locations that are returned to the user to select a venue to check-in at. Check-ins performed  by users are  forwarded  to the incorrect check-in detection module to handle noise in the check-in data. The user implicit feedback from the correct check-in operation is used to dynamically adjust the system parameters. We also presented novel approaches for  semantic labelling of the building floorplan and extending the  venues coverage of current LBSNs.
Extensive evaluation of CheckInside in four malls shows that CheckInside can infer the actual venue 99\% of the time within the top 5 venues in the candidate list.  In addition, it increases the coverage of current LBSNs by  37\%  by predicating names  of uncovered venues.

\section{ACKNOWLEDGMENTS}
This work is supported by a Google Research Award.

\begin{IEEEbiography}[{\includegraphics[width=1in,height=2in,clip,keepaspectratio]{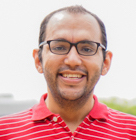}}]{Moustafa Elhamshary}   is a PhD student at Egypt-Japan University of science and Technology (E-JUST), Egypt. Currently, he is a research fellow at  Graduate School of Information Science and Technology, Osaka University, Japan. His research interests include location determination technologies,  mobile sensing, human activity recognition and location-based social networks.
\end{IEEEbiography}
\begin{IEEEbiography}[{\includegraphics[width=1in,height=1.25in,clip,keepaspectratio]{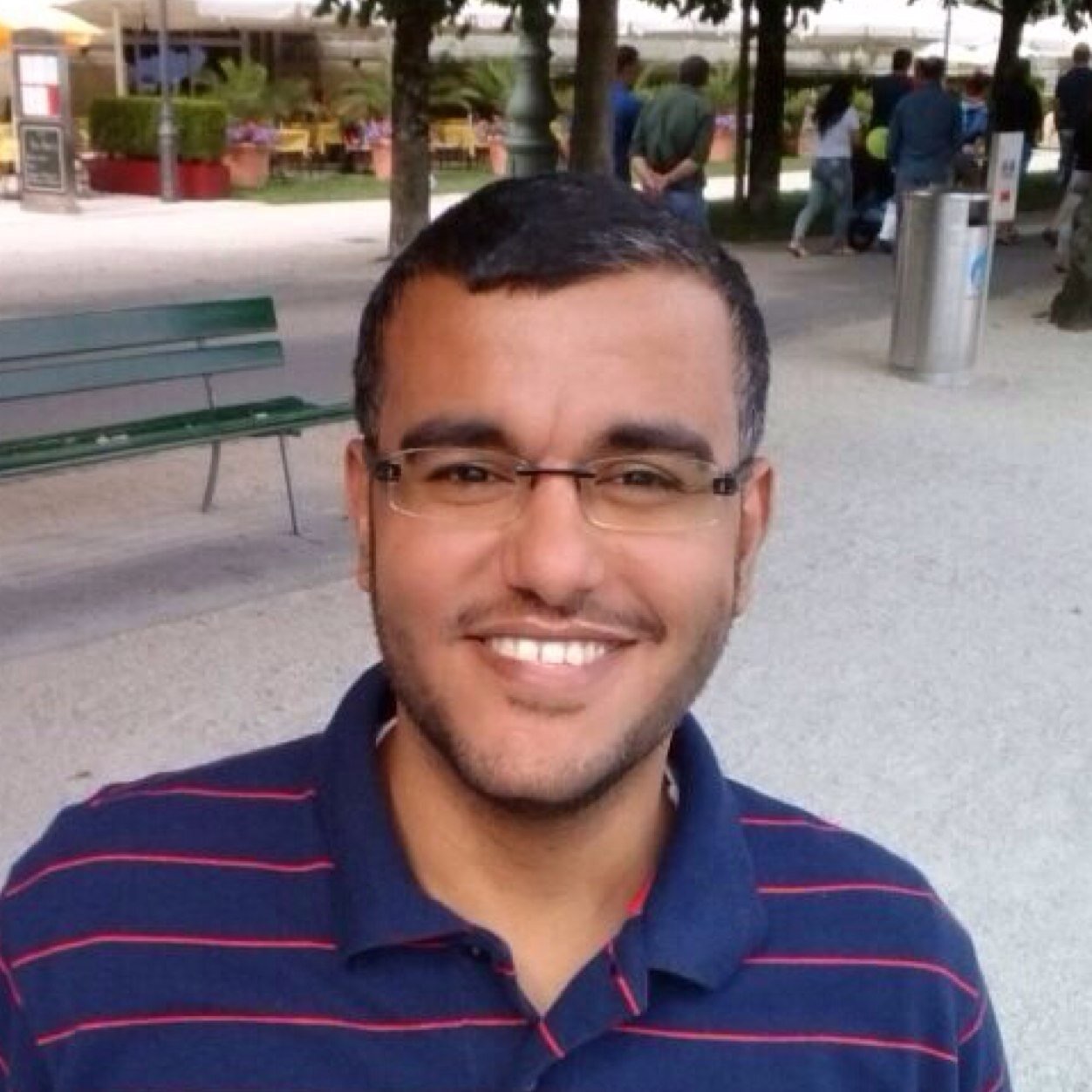}}]{Anas Basalamah} is an Assistant Professor at the Computer Engineering Department of Umm Al Qura University, Saudi Arabia.  He is a cofounder and director of the Wadi Makkah Technology Innovation Center at Umm Al-Qura University. He did his MSc and PhD Degrees at Waseda University, Tokyo in 2006, 2009 respectively. He worked as a Post Doctoral Researcher at the University of Tokyo and the University of Minnestoa in 2010, 2011 respectively.  He cofounded two tech startups: Navibees and Averos. His areas of interest include embedded networked sensing, ubiquitous computing,  participatory and urban sensing.
\end{IEEEbiography}

\begin{IEEEbiography}[{\includegraphics[width=1in,height=1.25in,clip,keepaspectratio]{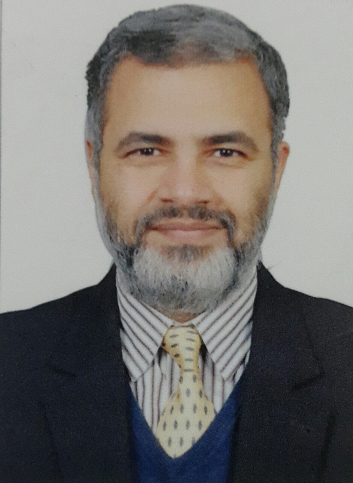}}]{Moustafa~Youssef}
is an Associate Professor at
Alexandria University and Egypt-Japan University
of Science and Technology (E-JUST),
Egypt. His research
interests include mobile wireless networks,
mobile computing, location determination technologies,
pervasive computing, and network
security. He has fifteen issued and pending
patents. He is an associate editor for the ACM TSAS, a previous area editor of the ACM MC2R, and served on the organizing
and technical committees of numerous prestigious conferences. Prof. Youssef is
the recipient of the 2003 University of Maryland Invention of the Year
award, the 2010 TWAS-AAS-Microsoft Award for Young Scientists, the 2012 Egyptian State
Award, the 2013 and 2014 COMESA Innovation Award, multiple Google Research Awards, the 2013 ACM SIGSpatial GIS Conference Best Paper Award, among others. He is also an ACM Distinguished Scientist.
\end{IEEEbiography}

\end{document}